\documentclass[12pt]{article}

\usepackage{amsmath,amsfonts,epsfig,color,latexsym,graphics}

\newcommand{\be}{\begin{equation}}
\newcommand{\ee}{\end{equation}}
\newcommand{\bea}{\begin{eqnarray}}
\newcommand{\eea}{\end{eqnarray}}

\let\newsection=\section
\renewcommand{\section}{\setcounter{equation}{0}\newsection}

\begin{document}

\begin{flushright}
TIT/HEP-577\\
November 2007
\end{flushright}
\vskip.5in

\begin{center}

{\LARGE\bf Gluon scattering in ${\cal N}=4$ Super Yang-Mills at finite temperature}
\vskip 1in
\centerline{\Large Katsushi Ito$^1$, 
Horatiu Nastase$^2$ and Koh Iwasaki$^1$}
\vskip .5in

\end{center}

\centerline{\large $^1$Department of Physics, Tokyo Institute of Technology}
\centerline{\large Ookayama 2-12-1, Meguro, Tokyo 152-8551, Japan}

\vskip .5in

\centerline{\large $^2$Global Edge Institute, Tokyo Institute of Technology}
\centerline{\large Ookayama 2-12-1, Meguro, Tokyo 152-8550, Japan}

\vskip .5in

\begin{abstract}
{\large
We extend the AdS/CFT prescription of Alday and Maldacena to finite temperature $T$, defining 
an amplitude for gluon scattering in ${\cal N}=4$ Super Yang-Mills at strong coupling from string theory. It is 
defined by a lightlike ''Wilson loop'' living at the horizon of the T-dual to the black hole in AdS space. Unlike 
the zero temperature case, this is different from the Wilson loop contour
 defined at the boundary of the AdS black hole metric, thus at nonzero $T$
there is no relation between gluon scattering amplitudes and the Wilson loop. We calculate 
a gauge theory observable that can be interpreted as the amplitude at strong coupling
for forward scattering of a low energy gluon ($E<T$) off a high energy gluon ($E\gg T$) in both 
cut-off and generalized dimensional regularization. The generalized dimensional regularization is defined in 
string theory as an IR modified dimensional reduction. For this calculation, the corresponding usual Wilson 
loop of the same boundary shape was argued to be 
related to the jet quenching parameter of the finite temperature ${\cal N}=4$ SYM plasma, while the 
gluon scattering amplitude is related to the viscosity coefficient.}

\end{abstract}

\newpage

\section{Introduction}

The AdS/CFT correspondence \cite{malda} is an important tool for calculating nonperturbative aspects of quantum field 
theories. Correlation functions of gauge invariant operators were calculated in ${\cal N}=4$ Super Yang-Mills,
and various other quantities were calculated in supersymmetric and nonsupersymmetric examples. For instance, 
for ${\cal N}=4$ SYM at finite temperature, when supersymmetry and conformal invariance are broken, 
the quark-antiquark potential was calculated \cite{bisy,bisy2} (following the $T=0$ calculation in
\cite{maldacena2,ry})
from a Wilson loop living in the gravity dual, a limit of the AdS black hole. However, the calculation of 
correlation functions or S matrices, relevant for physical scattering of particles, remains difficult for 
nonconformal and nonsupersymmetric theories. 

In a remarkable development, Alday and Maldacena \cite{am} proposed a way to calculate amplitudes for 
scattering of gluons (colored states) in ${\cal N}=4$ SYM at strong coupling, by making use of the factorization 
property of gluon amplitudes, ${\cal A}_n={\cal A}_{n,tree}a(p_i)$. Here $a(p_i)$ is a scalar function, given 
as an exponent of the string action for a worldsheet that ends on a lightlike polygon $C$, situated at the UV 
boundary of the metric T dual to $AdS_5$, $a(p_i)\sim e^{iS_{string}(C)}$. 
T duality was used just as a tool to make the calculation easier, as one starts with a Gross-Mende-type \cite{gm}
worldsheet ending on vertex operators for the gauge theory momenta, defined in the IR of the $AdS_5$. 
Because $AdS_5$ is the same as its T dual metric, $a(p_i)$ has the same value as the ${\cal N}=4$ SYM Wilson loop 
ending on the same lightlike polygon. This duality was found to hold also at weak coupling in \cite{dks,bht}. 

The calculation of the 4-point function done in \cite{am} (using the Wilson loop cusp found in \cite{krucz})
found that the result matches a conjectured result of Bern,Dixon and Smirnov (BDS) \cite{bds} (see also
\cite{abdk}) for the amplitude at any value of the coupling. The BDS conjecture gives a formula for any 
$n$-point amplitude, however there seems to be a discrepancy with the strong coupling 
calculation at large value of $n$ \cite{am3}. It was explained in \cite{dhks,am3} that 4- and 5-point amplitudes 
are fixed by conformal invariance, and a test of the functional form of the BDS conjecture appears only from 
6-point amplitudes and higher. In \cite{adin} a dual of a certain 6-point gluon amplitude was calculated, and it 
differs from the BDS conjecture, but it was argued that at least part of the discrepancy is due to the fact that 
the dual 6-point amplitude misses certain field theory diagrams. Other papers dealing with various aspects of 
the Alday-Maldacena proposal are \cite{bgk,am2,afk,ns,mmt,buch,ms,others}.

One would like however to define the same Alday-Maldacena duality for a theory that is closer to QCD, in order 
to be able to apply these methods to physical scattering of particles. One such possibility is to put the 
previous system at finite temperature, which would give a nonconformal and nonsupersymmetric theory.
Indeed, at finite temperature, the ${\cal N}=4$ SYM plasma is found to have properties quite similar 
to the QCD plasma \cite{nastase,gubser2}. In particular, the calculation of jet quenching in ${\cal N}=4$
SYM at finite $T$ \cite{gubser,hkkky,ct,lrw} seems to be applicable to QCD as well. It is  remarkable 
that the jet quenching parameter was in fact argued to be calculated using (partially) lightlike Wilson loops 
\cite{lrw,lrw2,lrw3}. There seems to be certain debate over this relation however \cite{aev}.
The quenching of gluons, relevant to our case, was calculated in \cite{cg}
(which also contains a detailed reference list on the subject).
 
As was already mentioned, in the finite temperature case, the usual Wilson loop calculation for the quark 
antiquark potential $V_{q\bar{q}}(L)$ was done already. However, at nonzero temperature the gravity dual and 
its T dual are no longer the same, so there is no relation between the Wilson loop and the gluon scattering 
anymore. In this paper we will analyze finite temperature gluon amplitudes, and we will define an extension 
of the Alday-Maldacena proposal to this case. Note that in the particular case we will analyze one of the gluons
will have energy $E\gg T$, thus can be treated as an external particle propagating through the plasma. 
In the process we will need to generalize the dimensional reduction 
procedure in the gravity dual, to account for the IR divergencies. We will find that Wilson loop-type 
observables in the T dual can be interpreted as the amplitudes for scattering of external gluons at finite temperature.

After a review of the zero temperature Alday-Maldacena proposal in section 2, we will look at possible Wilson 
loop-type calculations in the gravity dual, i.e. string worldsheets ending on lightlike polygons, 
in section 3. The polygon taken as an example
is a lightlike version of the usual contour taken in the calculation of $V_{q\bar{q}}(L)$ through the 
Wilson loop \cite{maldacena2,ry,bisy,bisy2}, 
which boosted gives the contour used for the jet-quenching parameter calculation \cite{lrw}.
In section 4 we will define the finite temperature gluon scattering and interpret the results. We will also show 
that the definition of the gluon scattering is consistent with having the string end on the boundary of the 
gravity dual, through an analysis of the Penrose diagram of the gravity dual.
In section 5 we will calculate the 4-point gluon scattering arising from the above lightlike polygon.
The Appendix contains a review of the extraction of $V_{q\bar{q}}(L)$ from the dual calculation of the Wilson 
loop and a calculation of the Penrose diagram of the gravity dual.

\section{Review of the Alday-Maldacena proposal at $T=0$}

Since there is no gauge group in AdS space,
the calculation of the AdS-CFT dual of gluon (colored states) amplitudes is made possible by
the property of ${\cal N}=4$ SYM gluon amplitudes with momenta $p_i$ that 
\be
{\cal A}_n={\cal A}_{n,tree}a(p_i)
\ee
where $a(p_i)$ is a scalar function and all color and polarization factors are in ${\cal A}_{n,tree}$. Thus one can 
obtain a result for $a(p_i)$ from $AdS_5$ that doesn't contain any gauge indices. The derivation of the proposal
starts with a Gross-Mende-type calculation \cite{gm}. One parallels the flat space calculation in the context 
of $AdS_5$, with Lorentzian signature metric
\be
ds^2=R^2\frac{dx_\mu dx^\mu+dz^2}{z^2}\label{ads}
\ee
where $x^{\mu}$ are the flat 4 dimensional Minkowski coordinates.
In AdS-CFT, the gluons scattered correspond to open strings, with a worldsheet ending on the IR of the $AdS_5$ metric
(\ref{ads}). More precisely, one considers an IR D-brane (at large $z=z_{IR}$), that can also act as an 
IR cut-off for the gluon amplitudes. In the high energy limit, the Gross-Mende type calculation for the string 
scattering of the open string states living on the D-brane is dominated by a classical worldsheet ending on the 
D-brane, with vertex operators corresponding to the states. However, the calculation is harder than in flat 
space, and it is computationally more convenient to make a "T duality" in the 3+1 flat dimensions $x^{\mu}$. This 
is a mathematical trick, since the 3+1 dimensional space is not compact. The transformation of coordinates is 
\be
\partial_{\alpha}y^{\mu}= i w^2(z)\epsilon_{\alpha\beta} \partial_{\beta } x^{\mu}
\ee
where $w(z)$ is the warp factor ($w(z)=R^2/z^2$ in AdS space). 
After this T duality that changes the coordinates as above and
inverts the metric as usual, the new metric is again $AdS_5$, in different coordinates
\be
ds^2=R^2\frac{dy_{\mu}dy^{\mu}+dr^2}{r^2};\;\;\;\; r\equiv\frac{R^2}{z}\label{dualads}
\ee
After the "T duality" transformation, the space is again noncompact, but the open string worldsheet has been changed.
Instead of vertex operators of momenta $k^{\mu}$ (Neumann boundary conditions), we have null segments 
\be
\Delta y^{\mu}=2\pi k^{\mu}
\label{deltay}
\ee
and the open string ends on a lightlike polygon formed by these segments (Dirichlet-type boundary condition). 
After the T duality, the D brane on which the string worldsheet ends is situated at small $r$, which is the 
UV boundary of the T dual AdS space (\ref{dualads}). Then the gluon amplitude is given by 
\be
a(p_i)=e^{iS_{string}(C)}
\ee
with $S_{string}(C)$ the action of the string ending on the lightlike polygon $C$. 
This is formally the same calculation 
as for the Wilson loop defined by $C$, but now in the T dual metric (\ref{dualads}). However since the 
T dual metric is again AdS, this means that the ${\cal N}=4$ SYM gluon amplitudes are related to the lightlike 
Wilson loop. This duality has been confirmed at weak coupling in \cite{dks,bht}. 

One puzzle that seems to arise is that in the original $AdS_5$, the ends of the open strings are defined in the 
IR, at $z\rightarrow \infty$, whereas one wants to define all physical quantities at the boundary of $AdS_5$, 
which is at $z=0$. The resolution of this aparent puzzle was done in \cite{am} by noticing that the boundary 
of the open string in T dual $y^{\mu}$ coordinates corresponds to $x^{\mu}\rightarrow\infty$
in the original $AdS_5$ coordinates. Together with 
$z\rightarrow \infty$ (such that all ratios are constant) this actually gives a point on the boundary of 
the original $AdS_5$, 
as a careful investigation of the $AdS_5$ Penrose diagram can show. In fact, the easiest way to see this is by 
finding the solution in global embedding coordinates for the original $AdS_5$, $X_A$, for which the boundary 
satisfies
\be
-X_{-1}^2-X_0^2+X_1^2+...+X_4^2=0
\ee
quotiented by overall rescalings. But one can also just analyze the Penrose diagram of the original $AdS_5$, 
and analyze where the string worldsheet ends. 

In \cite{am} a string worldsheet corresponding to a general 4-point gluon amplitude was analyzed. The 
lightlike polygon on the boundary is defined by $y_0,y_1,y_2$, whereas $y_3=0$. Then the string worldsheet 
extends in these directions together with $r$. By choosing a static gauge $y_1=u_1,y_2=u_2$
(where $u_1,u_2$ are worldsheet coordinates), 
the string action is 
\be
iS=-\frac{R^2}{2\pi\alpha '}\int dy_1dy_2\frac{\sqrt{1+(\partial_i r)^2-(\partial_i y_0)^2 -
(\partial_1r\partial_2 y_0-\partial_2 r\partial_1 y_0)^2}}{r^2}
\label{action}
\ee
whereas in a conformal gauge, the action is 
\be
iS=-\frac{R^2}{2\pi\alpha '} \int du_1du_2 \frac{1}{2}\frac{\partial r\partial r +\partial 
y_{\mu}\partial y^{\mu}}{r^2}
\label{confg}
\ee
For the solution $y_0=y_1y_2$, $r^2=(1-y_1^2)(1-y_2^2)$ (corresponding to the 4 gluon amplitude at $s=t$) 
the worldsheet has Euclidean signature, even though it lives in Lorentzian signature spacetime. 

The 4-gluon amplitude obtained from the above is IR divergent, as expected from general principles. There are 
2 possible ways to regulate the divergence, either by putting the IR D-brane at a finite value of $z_{IR}$
\cite{am,ms}, i.e. 
in the T dual the boundary of the open string is not at $r=0$, but at a finite $r_0$, or to use dimensional 
regularization. The first method corresponds to simple cut-off regularization in ${\cal N}=4$ SYM, whereas for the 
latter one uses dimensional reduction, i.e dimensionally reducing 10 dimensional Super Yang-Mills theory down to 
$d=4-2\epsilon$ dimensions. In the string dual this corresponds to having a D $p=3-2\epsilon$ brane in 10 dimensions
and going to its near horizon limit as usual. After the T duality this gives the metric
\be
ds^2= \sqrt{c_D\lambda_D}\left(\frac{dy^2_D+dr^2}{r^{2+\epsilon}}\right);\;\;\;
c_D=2^{4\epsilon}\pi^{3\epsilon}\Gamma(2+\epsilon);\;\;\;
\lambda_D=\frac{\lambda\mu^{2\epsilon}}{(4\pi e^{-\gamma})^{\epsilon}}
\label{metric}
\ee

Finally, the result of the calculation of the 4-point amplitude reproduces the result conjectured by 
Bern, Dixon and Smirnov (BDS) \cite{bds}
\bea
&&a(p_i)= ({\cal A}_{div,s})^2({\cal A}_{div,t})^2 \exp \{ \frac{f(\lambda)}{8} 
\ln^2\frac{s}{t}+const. \} \nonumber\\&&=
\exp \{ -\frac{f(\lambda)}{8}(\ln ^2 \frac{\mu^2}{-s}+\ln^2\frac{\mu^2}{-t})
-\frac{g(\lambda)}{2} (\ln \frac{\mu^2}{-s}+\ln \frac{\mu^2}{-t})+\frac{f(\lambda)}{8}
\ln^2\frac{s}{t}+const.\}
\eea
for the finite ($ O(\epsilon^0)$) part of the amplitude, where $f(\lambda)$ and $g(\lambda)$ are functions 
defined at weak coupling. Their strong coupling values are found to be 
\be
f=\frac{\sqrt{\lambda}}{\pi};\;\;\;\; g=\frac{\sqrt{\lambda}}{2\pi}(1-\ln 2)
\ee

\section{Finite temperature "Wilson loops"}

It is known how to introduce finite temperature in AdS-CFT. One just needs to add a black hole 
with mass $M$ in $AdS_5$, and 
take a scaling limit for $M\rightarrow\infty$ \cite{witten}, or equivalently take the near horizon limit of a 
near extremal black $p$-brane \cite{bisy}.  The two ways are related by a finite rescaling of coordinates. We will 
use the near horizon near extremal black $p$-brane form, since it is easier to dimensionally regularize in the 
desired way.

Specifically, the Witten construction \cite{witten} starts with the $AdS_{n+1}$ black hole in global coordinates
$(\bar{t},\bar{r},\Omega_{n-1})$,
\be
ds^2=-(1+\frac{\bar{r}^2}{R^2}-\frac{a^{n-2}}{\bar{r}^{n-2}})d\bar{t}^2+\frac{d\bar{r}^2}{1+\frac{\bar{r}^2}{R^2}
-\frac{a^{n-2}}{\bar{r}^{n-2}}}+\bar{r}^2d\Omega_{n-1}^2
\ee
and after a rescaling by terms depending on $M$, the mass of this AdS-BH (related to $a$), 
and taking $M\rightarrow\infty$, the 1 in the metric dissappears, and the $d\Omega_{n-1}^2$ becomes 
flat, thus getting
\be
ds^2=-(\frac{\rho^2}{R^2}-\frac{R^{n-2}}{\rho^{n-2}})dt'^2+\frac{d\rho ^2}{
\frac{\rho^2}{R^2}-\frac{R^{n-2}}{\rho^{n-2}}}+\rho^2d\vec{x}_{(n-1)}^2
\ee
With the rescaling $r=\rho (TR)/\pi,t=t'\pi/(TR), \vec{y}=\vec{x}\pi/T$ and introducing
$r_0=TR^2/\pi$ we get 
\be
ds^2=\frac{r^2}{R^2}[-dt^2(1-\frac{r_0^n}{r^n})+d\vec{y}_{(n-1)}^2]+R^2\frac{dr^2}{r^2(1-r_0^n/r^n)}\label{wittenm}
\ee
which is the same as the near horizon near-extremal D3-brane metric if $n=4$. 

We rewrite it for the physical $n=4$ case as (renaming the 4 dimensional coordinates $x_0,\vec{x}$)
\be
ds^2=\frac{R^2}{\tilde{r}^2}(-hdx_0^2+d\vec{x}^2+\frac{d\tilde{r}^2}{h})\label{metric2}
\ee
where 
\be
\tilde{r}=\frac{R^2}{r}; \;\;\; h=1-\frac{r_0^4}{r^4}=1-\frac{\tilde{r}^4}{\tilde{r}_0^4}
\ee

As we saw in the previous section, sometimes a "T duality" applied to this metric gives an easier description
of the physics. After the T duality, the metric is 
\be
ds^2=\frac{R^2}{r^2}(-\frac{dy_0^2}{h}+d\vec{y}^2+\frac{dr^2}{h})\label{tdualmetric}
\ee
where $h$ is the same and we use the $r$ coordinate instead of the $\tilde{r}$.

These two metrics are now different, unlike the $T=0$ case. But T duality does not change the physics, so this 
new metric should also describe AdS-CFT at finite temperature. But T duality relates complementary descriptions, 
so that different objects are easier to describe in one rather than in other. In particular, in the example 
of the previous section, Wilson loops were easier to describe in the original metric, whereas gluon amplitudes 
were easier to describe in the T dual metric. This point was somewhat obscured by the fact that the 
metrics were the same on both sides, resulting in a Wilson loop - gluon amplitude duality (also valid at small 
coupling). 

\begin{figure}[bthp]
\begin{center}\includegraphics{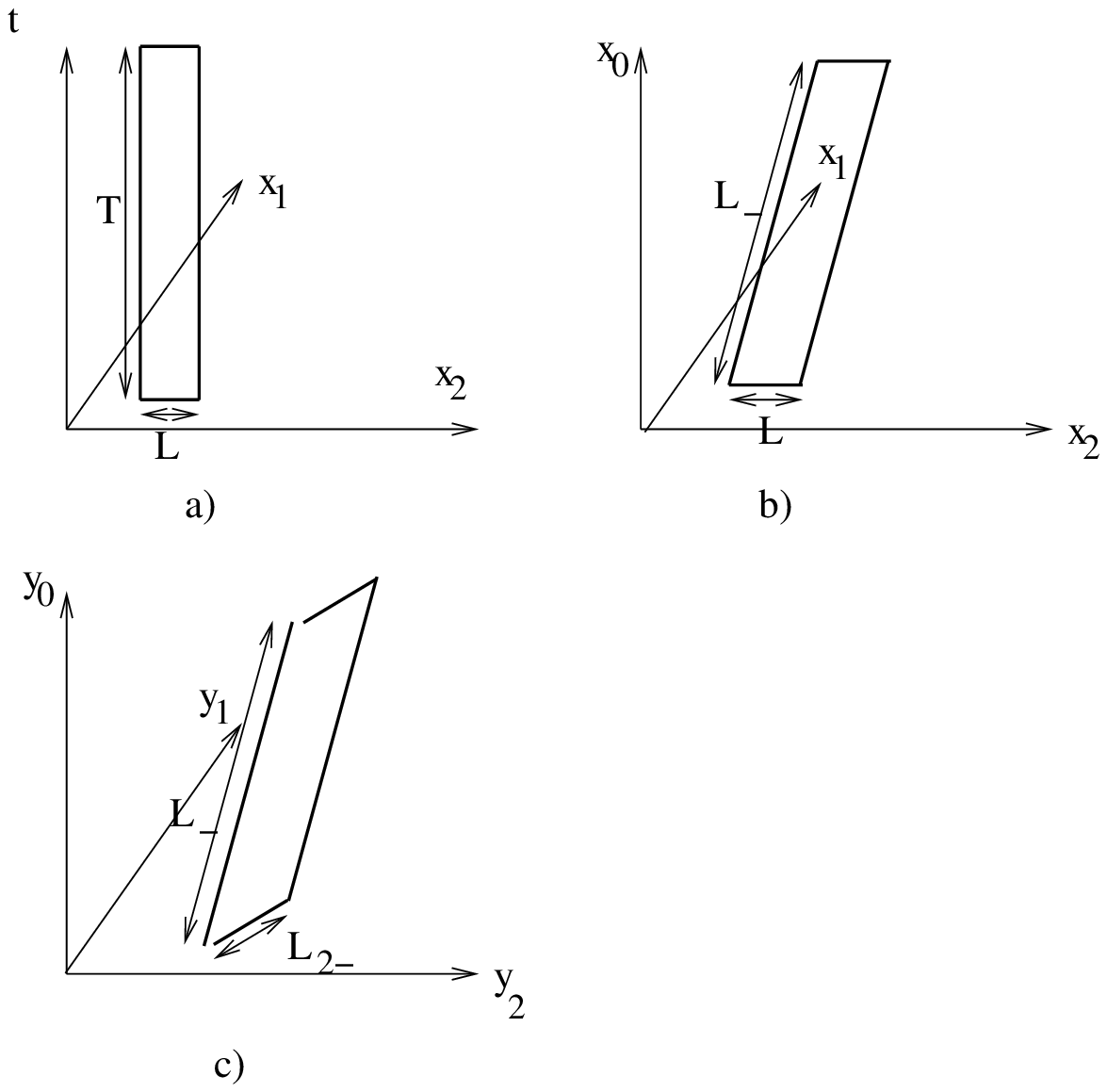}\end{center}
\caption{Contours for endpoints of worldsheets. a) Contour for the Wilson loop giving
$V_{q\bar{q}}(L)$, with $T\gg L$. b) Contour for the calculation of the Wilson loop 
giving the jet-quenching parameter, with $L_-\gg L$. c) Contour for the calculation of the gluon amplitude,
with $L_-\gg L_{2-}$. It is a lightlike polygon with rectangular projection on the spatial $(y_1,y_2)$
plane}\label{contours}
\end{figure}

Now however, the metrics are different, but we still should describe the same physics. We want to analyze string 
worldsheets that end on
Wilson loop contours in the original (\ref{metric2}) and T dual (\ref{tdualmetric}) metrics. These must correspond to 
different kinds of {\em observables} in ${\cal N}=4$ SYM.
The easiest to understand is the same kind of contour as the one used in the calculation of $V_{q\bar{q}}(L)$
(reviewed in the Appendix), namely a rectangle with one side much longer than the other, as in Fig.\ref{contours}a. 
Unlike the $V_{q\bar{q}}(L)$
case, where the long side is along the time direction and the short side is along a spatial direction, 
now we want both sides to be lightlike, as in Fig.\ref{contours}c.

For such a contour we can choose the static gauge as was done at $T=0$, namely $x_1=\tau, x_2=\sigma$, and obtain the 
analog of the action (\ref{action}). For the original AdS black hole metric (\ref{metric2}) we get the 
action
\be
iS=-\frac{1}{2\pi \alpha '}\int dx_1dx_2\frac{R^2}{\tilde{r}^2}\sqrt{1+\frac{(\partial_i\tilde{r})^2}{h}
-h(\partial_i x_0)^2
-(\partial_1\tilde{r}\partial_2x_0-\partial_2\tilde{r}\partial_1x_0)^2}
\label{adsbhact}
\ee
whereas for the T dual metric (\ref{tdualmetric}) with $y_1=\tau$, $y_2=\sigma$ we get 
\be
iS=-\frac{1}{2\pi\alpha '}\int dy_1dy_2\frac{R^2}{r^2}\sqrt{1+\frac{(\partial_ir)^2
-(\partial_i y_0)^2}{h}
-\frac{(\partial_1 r\partial_2y_0-\partial_2 r\partial_1y_0)^2}{h^2}}
\label{tdualact}
\ee

An intermediate case depicted in Fig.\ref{contours}b, with the long side lightlike, of length $L_-$ 
and the short side along a spatial direction, of length $L$, in the AdS black hole metric (\ref{metric2}), was 
argued in \cite{lrw} to be related to
 the jet-quenching parameter of the ${\cal N}=4$ SYM plasma at finite temperature. Indeed, 
it was known that the adjoint Wilson loop for this contour is related to the jet quenching parameter $\hat{q}$ by 
\be
<W^A(C)>=\exp\left[-\frac{\hat{q}}{4}\frac{L_-}{\sqrt{2}}L^2\right]
\ee
The authors of \cite{lrw} chose the gauge $\tau=x^-\equiv (x_0-x_1)/\sqrt{2}$, $\sigma=x_2$, and if $L_-\gg L$
one can assume that the string worldsheet is translationally invariant along the $\tau=x^-$ direction, therefore 
the string solution is only a function of $\sigma=x_2$. Specifically, the only nontrivial function is $r=r(\sigma)$,
for which they obtained the action (in $r$ coordinates)
\be
iS=-\frac{1}{2\pi \alpha'}\frac{L_-}{\sqrt{2}}\frac{r_0^2}{R^2} 2\int_0^{\frac{L}{2}}
d\sigma \sqrt{1+\frac{r'^2R^2}{fr^2}};\;\;\;\; f\equiv\frac{r^2}{R^2}(1-\frac{r_0^4}{r^4})\label{lrwact}
\ee
Its equation of motion is 
\be
r'^2=\gamma^2\frac{r^2f}{R^2}
\ee
where $\gamma$ is a constant,
thus we can see that the turning point $r'=0$ can only be at $f=0$, i.e. at the horizon $r=r_0$. That means that 
this Wilson loop is defined at the UV boundary of the space, but now matter how small $L$ is, the string drops all
the way to the horizon and then comes back to the boundary. 

It was argued in \cite{aev} that if one regularizes carefully the lightlike $v\rightarrow c$ limit of the spatial 
Wilson loop in Fig.\ref{contours}a, defining the Wilson loop at $r=\Lambda\gg r_0$, the leading contribution is 
not from the string dropping down to the horizon, but from string going up and down many times. Even if that 
is true, it is irrelevant for the main argument of this paper, as for the cases of interest the Wilson loop 
contours are defined at finite $r$ anyway. 

We can obtain the same action (\ref{lrwact}) from our action (\ref{adsbhact}) by choosing a similar ansatz, namely 
$x_0=x_1=\tau$ (replacing $\tau=x^-$), $x_2=\sigma$ we have already chosen, and translational invariance along 
$\tau$ means $r=r(\sigma)=r(x_2)$. 

Notice that if we take the zero temperature limit $r_0\rightarrow 0$, the action (\ref{lrwact}) becomes zero. The 
same result is obtained by using the previous ansatz, $y_0=y_1=\tau, r=r(\sigma)=r(y_2)$ in the Alday-Maldacena 
action at $T=0$, (\ref{action}). Thus the $T\rightarrow 0$ limit of this calculation is singular, and one does not 
get anything. It is easy to understand why this is so, since a string in AdS space can stretch in a lightlike 
direction without costing any energy. But this is not true once we break conformal invariance anymore, so at 
finite temperature we can have a nontrivial string configuration. 

We now turn to the case of interest, where all sides of the Wilson loop contour are lightlike. With respect to 
the previous case, we will have a modification of the shape of the worldsheet near the corners of the contour, but 
because we work in the $L_-\gg L$, this modification is subleading in the action. More importantly, now the whole 
worldsheet is "tilted" since the $\sigma$ direction is not spacelike of length $L$ anymore, but lightlike of length 
$L_{2-}$. 

As a result, if we also choose an ansatz for a flat worldsheet (in the $y_0,y_1,y_2$ directions), which is appropriate
for the case $L_-\gg L_{2-}$  (translational invariance in the $\tau$ direction), then we have $y_0=\sigma+\tau
=y_1+y_2$ and $r=r(\sigma)=r(y_2)$. If we plug this ansatz in the $T=0$ action (\ref{action}) we obtain 
$S\propto \int d\sigma 1/r^2$, which again does not define $r(\sigma)$. 

Plugging the ansatz in the $T\neq 0$ action (\ref{adsbhact}) instead, we obtain (in $r$ coordinates)
\be
iS=-\frac{1}{2\pi \alpha '}\frac{L_-}{\sqrt{2}}\frac{r_0^2}{R^2}2\int_0^{L_{2-}/(2\sqrt{2})}
dx_2\sqrt{2-\frac{r^4}{r_0^4}+\frac{r'^2 R^2}{r^4(1-\frac{r_0^4}{r^4})}}
\ee

The Lagrangian density does not depend explicitly on $x_2$, so we can consider it a one dimensional mechanics 
problem with $x_2=$ "time", therefore the Hamiltonian for this motion will be a constant (independent of $x_2$), 
and equal to $\equiv 1/\alpha$. This gives the equation of motion
\be
r'^2=\frac{r^4-r_0^4}{R^4}\left\{ (2-\frac{r^4}{r_0^4})^2\alpha^2-2+\frac{r^4}{r_0^4} \right\}
=\frac{(2r_0^4-r^4)(r^4-r_0^4)\alpha^2}{r_0^4R^4}(2-\frac{1}{\alpha^2}-\frac{r^4}{r_0^4})
\ee

Now the turning point $r'=0$ is either at $r=r_0$, at $r=\bar{r}_1=(2-1/\alpha^2)^{1/4}r_0<2^{1/4}r_0$
or at $r_2=r_0 2^{1/4}$. Therefore $r'^2 \geq 0$ (so $r$ is real) if $r\geq r_2$ or $r_0\leq r\leq \bar{r}_1$. 
The second possibility appears only if $\alpha^2>1$, in which case $\bar{r}_1>r_0$.

The first case, $r\geq r_2$, with $r_2$ turning point, corresponds to the usual Wilson loop. The string worldsheet 
ends on the Wilson loop contour in the UV, at $r=\infty$, and it drops down to $r=r_2>r_0$, after which it goes back to 
infinity. We see that unlike the partially lightlike Wilson loop in \cite{lrw}, the string drops only to a finite 
distance away from the horizon, at $r_2=r_0 2^{1/4}$. But like in \cite{lrw}, the string drops to the same place 
($r_2$), independent of how small $L_{2-}$ is.

Now however we also have a most puzzling case, if $\alpha^2>1$ (Hamiltonian $<1$)
of a string worldsheet that can end on a Wilson loop contour defined 
at any point between the horizon $r_0$ and $r_1=(2-1/\alpha^2)^{1/4}r_0<2^{1/4}r_0$, and it can drop either to 
the horizon $r_0$ (where it turns back), OR upwards, to $r=r_2$, where it turns back. It is not clear 
what would be the possible physical significance of this string worldsheet in ${\cal N}=4$ SYM.
It could be that like in \cite{aev} this signals the possibility of a worldsheet with any number $n$ of turns, thus 
that in this case observables are hard to define. 

We now repeat the same analysis in the T dual AdS black hole metric (\ref{tdualmetric}). Plugging the same ansatz
$y_0=\sigma+\tau=y_1+y_2$ and $r=r(\sigma)=r(y_2)$ in the T dual action (\ref{tdualact}), we get 
\be
S=\frac{1}{2\pi \alpha '}\frac{L_-}{\sqrt{2}}2\int_0^{L_{2-}/(2\sqrt{2})}dy_2\frac{R^2r_0^2}{r^4-
r_0^4}\sqrt{r'^2+\frac{r^4+r_0^4}{r_0^4r^4}(r^4-r_0^4)}\label{tdact1}
\ee

Notice the factor of $i$ difference between the Wilson loop calculation (in the AdS black hole metric) and the
T dual Wilson loop (for the T dual AdS black hole metric). 
Again, the Lagrangian density is independent of $y_2$, thus the Hamiltonian (with $y_2=$ "time") is constant 
as a function of $y_2$, and is defined to be $1/\alpha$. The equation of motion is then
\be
r'^2=\frac{r^4+r_0^4}{r_0^4r^8}[R^4\alpha^2(r^4+r_0^4)-r^4(r^4-r_0^4)]\label{eom1}
\ee

The turning point $r'=0$ is then at 
\be
r=r_1=r_0\left[\frac{b+1}{2}+\sqrt{(\frac{b+1}{2})^2+b}\right]^{1/4};\;\;\; b\equiv \frac{\alpha^2 R^4}{r_0^4}
\label{turnpnt}
\ee
only (and not at the horizon $r_0$, where $r'^2(r_0>0$)
and $r$ is real ($r'^2\geq 0$) only if $r_0\leq r\leq r_1$. That means that now we have a single 
possible string worldsheet, one that ends on the Wilson loop contour at $r_0\leq r\leq r_1$, and "drops" to 
$r=r_1(\alpha)$, where is turns back. Note that in the T dual metric, the horizon $r=r_0$ is in the UV
(or at least where the UV was at $T=0$), so it does make sense to put the Wilson loop contour close to it.

This string worldsheet is clearly the analog of the string worldsheet giving the gluon scattering amplitude at $
T=0$. Note that $r'$ is now finite at the horizon $r=r_0$, however if we define the Wilson loop contour exactly 
at $r_0$, the integral for the action $S$ is divergent at the horizon. This is consistent with the fact that 
gluon amplitudes are IR divergent even at finite temperature. We will analyze these Wilson loops in more detail 
in the next section. 

Therefore there are 3 kinds of string worldsheets ending on lightlike Wilson loop contours at finite temperature.
We have the string worldsheet giving the finite temperature Wilson loop, defined in the AdS black hole metric 
(\ref{metric2}) and ending on a contour situated at the 
UV boundary ($r\rightarrow \infty$), that stretches down to $r=r_2$=fixed. This Wilson loop is in fact related 
with the partially lightlike Wilson loop of \cite{lrw} that we have described, by an infinite boost (tilting 
the spatial direction $L$ to the lightlike direction $L_{2-}$). The second possibility is the string worldsheet 
giving the gluon scattering amplitude, defined in the T dual AdS black hole metric (\ref{tdualmetric}), ending on 
a contour situated at the horizon and stretching to $r=r_1(\alpha)$. Finally, there is the strange possibility 
that the string worldsheet defined in the AdS black hole metric (\ref{metric2}) ends on a contour situated 
between $r_0$ and $\bar{r}_1$ and can stretch either to the horizon $r_0$ or to $r_1$.

All these 3 string worldsheets define different observables at finite temperature, which can be in principle 
defined in either the AdS black hole metric (\ref{metric2}) or its T dual (\ref{tdualmetric}).
In (\ref{metric2}) the Wilson loop is defined by the string worldsheet ending on the contour
at infinity, whereas gluon scattering 
is defined by a worldsheet with vertex operator insertions at the horizon (analogous to the discussion at $T=0$ 
in the previous section). In (\ref{tdualmetric}), gluon scattering is defined by the string ending on the contour
at the horizon, whereas the Wilson loop is defined by the string worldsheet with vertex operator insertions at 
$r\rightarrow\infty$ (in the IR of the T dual metric (\ref{tdualmetric})). It is not clear to what does the 
third string worldsheet correspond in either metric.

\section{Interpretation and gluon scattering}

Now that we have seen what kind of string worldsheets ending on lightlike contours
we have at finite temperature, we can define more precisely gluon scattering, corresponding to the worldsheet 
in the T dual AdS black hole metric (\ref{tdualmetric}). A question that we will not try to address is the definition 
of the external gluon states. For our purposes it is enough that we have a well defined observable in the gravity dual,
showing that such a definition should exist.

In ${\cal N}=4 $ SYM at finite temperature, it is not a priori obvious that we still have the factorization 
property of amplitudes
\be
{\cal A}_n={\cal A}_{n,tree}a(p_i)
\ee

However, from the string theory side of AdS-CFT, introducing finite temperature just changes the background of 
the gravity dual, but nothing else changes. So we expect the value of $a(p_i)$ to change, but not the factorization
property. It also means we will continue to have 
\be
a(p_i)=e^{iS_{string}(C)}
\ee
but now, since the T dual metric is different than original one, we will not have $a(p_i)=\hat{w}(p_i)$ anymore (where 
$\hat{w}(p_i)$ 
is the corresponding Wilson loop quantity). In fact, we have seen in the previous section that the worldsheets
corresponding to the Wilson loop and the gluon scattering are in fact different. This was to be expected, since 
the duality relation $a(p_i)=\hat{w}(p_i)$ was derived (even at small coupling) from a (dual) conformal symmetry 
\cite{dks,bht}, that is not present anymore at finite temperature.

Another difference is that the IR of the AdS black hole metric (\ref{metric2}) is not at $r=0$ anymore, but at 
the horizon $r=r_0$. Therefore this is where we must put the D-branes on which the open string ends, and IR 
regularization means now to put the D-branes at $r=r_0(1+\epsilon)$. The endpoints of the string 
correspond to vertex operators of given 4-momenta (4 dimensional Neumann boundary conditions), and as at $T=0$ it 
is easier to make the "T duality" transformation to the metric (\ref{tdualmetric}). After the T duality, the 
open string worldsheet ends on a lightlike polygon at $r=r_0$
with sides defined by 
\be
\Delta y^{\mu}=2\pi\alpha ' k^{\mu}
\ee
Here we have reintroduced $\alpha '$ to emphasize that this relation is independent on the metric (the "radius"
of the "compact" dimension). In fact, the above relation is a generalization of the usual exchange of momentum 
$n/R$ modes with winding $mR/\alpha '$ modes. Now $n/R=k^{\mu}$ and $m=\Delta \theta/(2\pi)$, thus $mR=\Delta 
y^{\mu}/(2\pi)$. The fact that the above relation is independent of the metric is important, since 
now in the metric (\ref{metric2}) $g_{00}\neq g_{ii}$, and we make a "T duality" transformation on all 4 dimensions. 
The above relation implies that the segments defining the Wilson loop contour are still lightlike.

However, the T duality transformation acts also on the coordinates by 
\be
\partial_{\alpha}y^{\mu}=iw^2(z)\epsilon_{\alpha\beta}\partial_{\beta}x^{\mu}
\ee
in (worldsheet) conformal gauge, and  the "warp factor" $w(z)$ is now different for $g_{00}$ and $g_{ii}$. 

This means that in the original $x^{\mu}$ (complex) coordinates, the lighlike polygon is not lightlike 
anymore (i.e. $\Delta x_i^2-\Delta t^2\neq 0$ now), unlike the $T=0$ case. However, these are 
complex coordinates, so they don't have much physical significance.
Note that in the T dual metric $r=r_0$ is in the UV and plays 
the role of "boundary" of space, thus it is OK to define the Wilson loop contour there.

But like at $T=0$ we seem to have the puzzle that in the original AdS black hole metric (\ref{metric2})
the endpoints of the open string are at the horizon (in the IR), instead of at the UV boundary at $r=\infty$.
We therefore need to analyze the endpoints of the open string in $x^{\mu}$ coordinates and see whether they are 
on the boundary of space. For that, we must see first if any of the $x^{\mu}$ go to infinity on the solution 
at $r\rightarrow r_0$. 



Since $\partial_1y_0=\partial_2y_0=1$, but $g_{00}\rightarrow 0$ at the horizon for the 
AdS-BH, it means that $\partial_\alpha t$ is infinite at the horizon, thus $t=+\infty$ at the horizon for 
the open string solution  in the T dual metric.
On the other hand, the rest of $\partial_\alpha y^i$ are also finite, but $g_{ii}$ is 
finite at the horizon, so $\partial_\alpha x^i$ is finite at the horizon.
Therefore  the boundary of our solution in the original coordinates is $t=+\infty, x^i$=finite, 
$r=r_0$.

We have analyzed the Penrose diagram of the Poincar\'{e} AdS black hole in the Appendix. It was necessary to do it, 
since the metric (\ref{metric2}) is a limit of the usual AdS black hole, and we also wanted to 
find the explicit coordinate transformations needed. We have proven there that the boundary in fact touches the 
horizon $r=r_0$ on the limit $t=\pm \infty$, $x_i$ arbitrary, therefore the endpoints of the open string in $x^{\mu}$
coordinates are indeed on the  boundary of the gravity dual. 

One might be also puzzled that we took $t\rightarrow\infty$, but we had a T duality on the time direction as 
well. However as we mentioned, the "T duality" transformation does not mean that either the original or the final 
space is compact. Even though we are at finite temperature, we work in the Lorentzian section, so time is 
not periodic. 

We now turn to understanding the gluon scattering amplitudes at finite temperature. At finite temperature, Lorentz 
invarariance is broken by the presence of the heat bath (which defines a preferred reference frame). That means 
that the finite temperature amplitudes can depend on more parameters than at $T=0$. 

For the 4-point amplitude, at $T=0$ we have only 2 parameters, the Mandelstam variables $s$ and $t$. 
At finite $T$ however we have 5 parameters: 3 independent
on-shell momenta (3 components) give 9 parameters, minus 3 rotations, minus one constraint
(= the on-shell condition for the sum of the 3 momenta) gives 5. 
They are: the 3 energies of the independent 4-momenta, and 3 angles between them, minus one constraint. Normally, there would also be 3 boosts, which would bring it down to 2 parameters, s and t. But now those change the 
physics.

At finite temperature $T$, photons acquire a mass $eT$, and gluons aquire also a mass of order $gT$: the 
propagators get renormalized by thermal loops and one can interpret them as a thermal mass. Since there is a 
mass, one would be inclined to think that the IR divergencies would dissappear, however that is not the case. 
Gluon scattering amplitudes are also in general IR divergent. This fits with the observation made in the previous 
section that the string action with boundary at the horizon in the T dual metric (\ref{tdualmetric}) is IR divergent.

The particular string worldsheet that we described in the last section, ending on a contour with $L_-\gg L_{2-}$ and 
a planar topology, corresponds to a 4-point amplitude with the color ordering such that $y_0$ increases over two 
adjacent sides, and then decreases over the next. That means that the null external momenta 
$k_1,k_2$ are incoming, and $k_3,k_4$ are outgoing. Also, $k_1=k_3$ has length $L_-$ and $k_2=k_4$ has length $L_{2-}$.
This is then forward scattering of an energetic gluon off a soft gluon ($L_-\gg L_{2-}$).

For such a process, the Mandelstam variables are
\bea
&&-s=(k_1+k_2)^2=\frac{1}{(2\pi\alpha ')^2}[\frac{L_-^2+L_{2,-}^2}{2}-\frac{(L_{2-}+L_-)^2}{2}]=-\frac{
L_{2-}L_-}{(2\pi\alpha ')^2};\nonumber\\
&&-t=(k_2+k_3)^2=\frac{1}{(2\pi\alpha ')^2}[\frac{L_-^2+L_{2,-}^2}{2}-\frac{(L_{2-}-L_-)^2}{2}]=\frac{L_{2-}L_-
}{(2\pi\alpha ')^2}=+s;\nonumber\\
&&u=0
\eea

As we said, Lorentz invariance is broken by the heat bath, but at $T=0$ $s$ and $t$ still parametrize the 
amplitude. It is therefore instructive to calculate the $T=0$ value of the 4-point amplitude given by the 
BDS formula, tested by Alday and Maldacena for the color ordering in, out, in, out,  for these invariants
\bea
&&a_4(s,t)=\exp(-\frac{f(\lambda)}{4}\ln^2\frac{L_{2-}/\alpha 'L_-/\alpha '}{(2\pi\mu)^2}-\frac{g(\lambda)}{2}\ln 
\frac{L_-/\alpha 'L_{2-}/\alpha '}{(2\pi\mu)^2})\nonumber\\&&\times
\exp(-\frac{i\pi}{4}(f(\lambda)\ln\frac{L_-/\alpha 'L_{2-}/\alpha '}{(2\pi \mu)^2}+2g(\lambda)+const.))\label{amplres}
\eea

This result contains both real and imaginary parts, and thus clearly could not be obtained from a string worldsheet
action, since the worldsheet has either Euclidean or Lorentzian signature, but not both. 
This property is confirmed by the fact that the Alday-Maldacena action (\ref{action}) is zero on the 
lightlike ansatz, as we noticed. This was due to the fact that in $AdS$ space we can stretch a string along a null 
direction at no cost, but this is no longer true if we break the conformal invariance. We note that the above formula
is just an application of the BDS conjecture to the in, in, out, out ordering, but otherwise it was not directly 
checked in any computation.

The result (\ref{amplres}) means then that we cannot obtain the $T=0$ case as a limit of the finite $T$ calculation,
so there should be an obstruction to taking this limit. This is indeed what we will find in the next section, where 
we will see that $L_{2-}/\alpha '<T$, thus the limit does not make sense.

At $T=0$, another way to understand the fact that the AdS string action is zero is to change the colour ordering, 
namely to consider $k_1$ and $k_3$ incoming and $k_2$ and $k_4$ outgoing (and $k_1=k_2$, $k_3=k_4$). In this case, 
the string worldsheet collapses to two lines, so gives a zero action. Since different colour orderings are 
supposed to give equivalent ways to calculate the same $a(p_i)$, it shows that we can not do it. 

The same argument at $T\neq 0$ does not work due to the presence of the heat bath. The forward scattering with 
$k_1$ and $k_3$ incoming and $k_2$ and $k_4$ outgoing colour ordering still gives a collapsed string worldsheet, 
but the ordering we chose doesn't.

The 4-point amplitude we are interested is not the most general one. As we mentioned, we should have 5 parameters
(2 invariants and 3 boosts), but we have only 2 now, $L_-$ and $L_{2-}$, moreover with one condition between them, 
that $L_{-}\gg L_{2-}$. That means that we cannot extract the most general 4-point gluon amplitude. 

This is unfortunate, since if we did, we could calculate the viscosity coefficient at finite temperature. 
Indeed, one has the relation \cite{lebellac}
\bea
&&\eta=\frac{\beta}{4}\nu_g^2\int d\tilde{p}_1d\tilde{p}_2d\tilde{p}_3d\tilde{p}_4
n_1n_2(1+n_3)(1+n_4)\nonumber\\&&
(2\pi)^4\delta^{(4)}(p_1+p_2-p_3-p_4)\bar{|{\cal M}|^2}[\Phi_1+\Phi_2-\Phi_3-\Phi_4]^2\label{visco}
\eea
where $\nu_g=16$ is the sum over colours and spins on the out state, $\bar{|{\cal M}|^2}$
is the gluon scattering amplitude squared summed over spins and colours in final state and 
averaged over the same in the initial state (thus is related to the Wilson loop calculation)
and $n$ and $\Phi$ characterize a solution of the Boltzmann equation.

\section{Calculation of a finite temperature gluon amplitude}

We now turn to the calculation of the 4-point gluon amplitude we described. We have a Wilson loop in the 
metric (\ref{tdualmetric}) that ends at the horizon $r=r_0$ on the lightlike contour C in fig.\ref{contours}c.
The action is given by (\ref{tdact1}), with equations of motion (\ref{eom1}) and turning point (\ref{turnpnt}).
Substituting the equations of motion in the action we get 
\be
S=\frac{\alpha R^4}{2\pi \alpha '}\sqrt{2}L_-\int_0^{\frac{L_{2-}}{2\sqrt{2}}}dy_2
\frac{r^4+r_0^4}{r^4(r^4-r_0^4)}
\ee

We define $z=r/r_0$ and $a=r_1/r_0$ and using the equations of motion for $r'=dr/dy_2$ we replace $dy_2$ by $dz$. 
Also replacing $\alpha$ from 
\be
a=\left[\frac{b+1}{2}+\sqrt{(\frac{b+1}{2})^2+b}\right]^{1/4};\;\;\; b=\frac{\alpha^2R^4}{r_0^4}\label{turnpnt2}
\ee
we get
\be
S=\frac{R^2}{2\pi \alpha '}\frac{\sqrt{2}L_-}{r_0}\frac{1}{\sqrt{a^4+1}}
\int_1^a\frac{dz}{z^4-1}\sqrt{\frac{z^4+1}{\frac{z^4+1}{a^4+1}-\frac{z^4}{a^4}
\frac{z^4-1}{a^4-1}}}\label{tdact2}
\ee

Here $a=a(L_{2-}/r_0)$ and is defined by integrating $y_2$ as 
\be
\frac{L_{2-}}{2\sqrt{2}}=\int dy_2=\int \frac{r_0dz}{r'}=\frac{r_0}{a^2\sqrt{a^4-1}}\int_1^{a}
\frac{dz z^4}{\sqrt{z^4+1}\sqrt{\frac{z^4+1}{a^4+1}-\frac{z^4}{a^4}\frac{z^4-1}{a^4-1}}}
\ee

The action (\ref{tdact2}) is log divergent at the horizon $a=1$ ($r=r_0$): $\sim \int_1 dz/(z-1)$, 
but finite at $z=a$. That means that we need to introduce a regularization that will change the 
behaviour of the action at the horizon. 

\begin{figure}[bthp]
\begin{center}\includegraphics{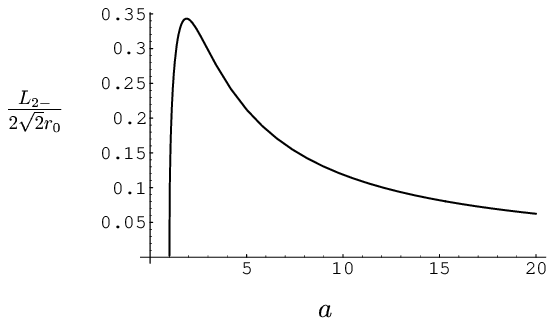}\end{center}
\caption{$L_{2-}/r_0$ as a function of $a$}\label{graph1}
\end{figure}

\begin{figure}[bthp]
\begin{center}\includegraphics{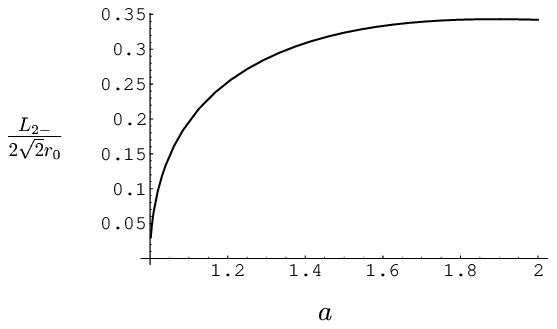}\end{center}
\caption{$L_{2-}/r_0$ as a function of $a$}\label{graph2}
\end{figure}

The integral for $L_{2-}$ is finite however, which means that we can calculate it without the need for 
regularization. We could not find an analytical solution, but the result of numerical integration for 
$L_{2-}/r_0$ as a function of $a$ is shown in Fig.\ref{graph1} and \ref{graph2}.


From (\ref{turnpnt2}), as $\alpha $ varies between 0 and $\infty$, $a$ varies 
between 1 and $\infty$. From the graph, then $L_{2-}$ varies between 0 and 0, with a maximum of about 
$L_{2-}^{max}/r_0\simeq 0.35$ at  $a_{max}\simeq 1.8911$.

However,  that means that only one branch of the $L_{2-}(a)$ graph is physical ($1<a<a_{max}$ or $a_{max}<a<\infty$).
Which one should be decided by the physical condition that $S(L_{2-}=0)=0$, which will be satisfied by either 
$S(a=1)$ or $S(a=\infty)$. It seems natural to guess that $a=1$ is the correct choice, but from (\ref{tdact2}) 
it is not a priori obvious, so we need to wait for the regularization. 

Therefore we need $L_{2-}<L_{2-}^{max}\simeq 0.35 r_0$ for this solution to exist. This in particular implies 
that the $T\rightarrow 0$ ($r_0\rightarrow 0$) limit of this calculation does not exist. This is good, since 
we saw that at $T=0$ we could not get such a solution, and the BDS formula gave a result with both real 
and imaginary parts, which cannot be obtained from a string worldsheet. 

We now need to regularize the  integral in (\ref{tdact2}). 

{\bf Cut-off regularization}

The simplest choice is cut-off regularization, i.e. defining the contour $C$ not at $r=r_0$, but at 
$r=r_0(1+\epsilon)=r_0+\Lambda$ (the IR D-brane on which we define the scattering amplitude is 
situated a bit away from the horizon). This will correspond to a simple IR cut-off in the CFT.
We then integrate (\ref{tdact2}) only down to $r=r_0(1+\epsilon)$, thus $z=1+\epsilon$. 

This is the same kind of regulator commonly used in Wilson loop calculations. 
The reason that Alday and Maldacena did not 
investigate this cut-off is twofold: they did not have a general solution for a Wilson loop 
defined a bit away from the IR, and the field theory results were in dimensional regularization.
None of these arguments is relevant for us: we do have a solution if we define the Wilson loop 
away from the horizon, and there is no field theory prediction to test against.

\begin{figure}[bthp]
\begin{center}\includegraphics{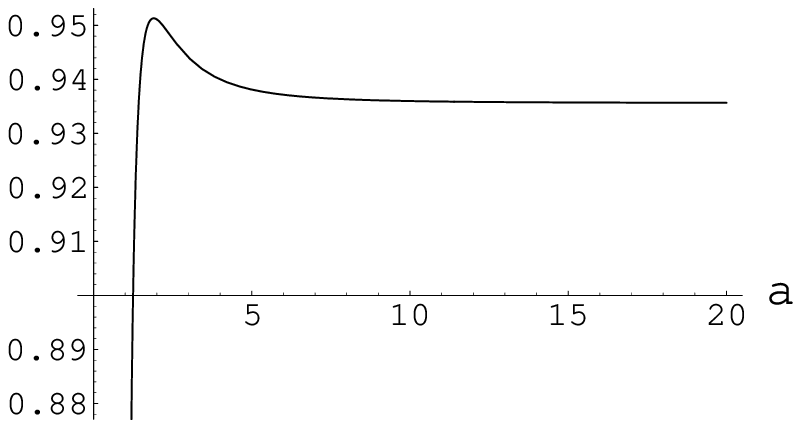}\end{center}
\caption{Integral in S, divided by $\sqrt{a^4+1}$, as a function of $a$, 
for $\epsilon=0.01$}\label{graph5}
\end{figure}

\begin{figure}[bthp]
\begin{center}\includegraphics{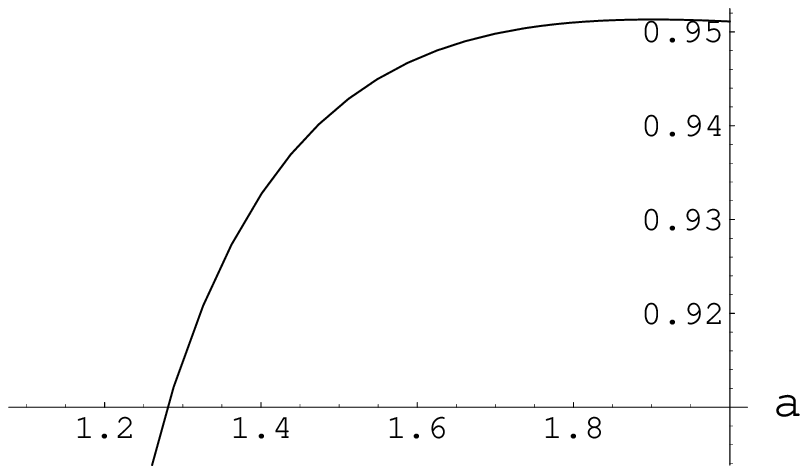}\end{center}
\caption{Integral in S, divided by $\sqrt{a^4+1}$, as a function of $a$, 
for $\epsilon=0.01$}\label{graph6}
\end{figure}

Integrating (\ref{tdact2}) only down to $z=1+\epsilon$ we get a finite result, but cannot find an 
analytic form for the result. Taking $\epsilon =0.01$ and evaluating numerically $S$, we obtain 
(up to constants), the graph in 
Fig.\ref{graph5} and Fig.\ref{graph6}.


Therefore the integral goes to a constant at $a\rightarrow
\infty$, where $L_{2-}\rightarrow 0$. But this is not the correct behaviour, as we argued. So the 
correct branch of $L_{2-}(a)$ is in fact the lower one ($1<a<a_{max}\simeq 1.9$). For $a\rightarrow 1+
\epsilon$, the action does go to zero, and $L_{2-}\rightarrow 0$ as well.
The maximum of $L_{2-}$ and the maximum of $S$ seem to be numerically at about the same $a_{max}$.  

We cannot calculate analytic results for $S$ and $L_{2-}$
at general values of $L_{2-}$, therefore general values of $a$, but we can calculate the small $L_{2-}$ 
asymptotics, i.e. $a\rightarrow 1+\epsilon$. Defining $a-1-\epsilon
\equiv \delta\rightarrow 0$, the action becomes 
\be
S\simeq \left[\frac{R^2}{2\pi \alpha '}\frac{\sqrt{2}L_-}{r_0}\right]
\frac{\sqrt{\delta}}{2\sqrt{\delta+\epsilon}}\sinh^{-1}(\sqrt{\frac{\delta}{\epsilon}})
\ee

If we have $\epsilon/\delta \ll 1$, i.e. if we take first $\epsilon\rightarrow 0$ and then $\delta\rightarrow 0$, as 
we should, then we obtain 
\be
S\simeq \left[\frac{R^2}{2\pi \alpha '}\frac{\sqrt{2}L_-}{r_0}\right] \frac{1}{2}
\ln 2\sqrt{\frac{\delta}{
\epsilon}}
\ee

Then we can also calculate $L_{2-}(a)$ and we get
\be
L_{2-}\simeq 2 r_0\sqrt{\delta}\Rightarrow \delta\simeq \left(\frac{L_{2-}}{2r_0}\right)^2
\ee

Replacing this into the action we obtain
\be
S\simeq \left[\frac{R^2}{2\pi \alpha '}\frac{\sqrt{2}L_-}{r_0}\right]\times \frac{1}{4}\ln\left[\left(\frac{
L_{2-}}{r_0}\right)^2\frac{1}{\epsilon}\right]
\ee
where $\epsilon=\Lambda/r_0$. This result is only the leading term in the action, but since 
the divergence is $o(\ln \epsilon)$, the expansion in epsilon will give extra terms of order $\epsilon\ln\epsilon\rightarrow 0$, as we can easily check. 
The expansion in $\delta$ and $\eta=z-1-\epsilon$ 
will give terms with an extra $\delta$, that are subleading at $\delta\rightarrow
0$. We argue below that in fact $\delta \ln\epsilon$ terms should also cancel. 

Thus the full finite part of the integral
at small $L_{2-}$ is
\be
S_{finite}=\left[\frac{R^2}{2\pi \alpha '}\frac{\sqrt{2}L_-}{r_0}\right]\times \frac{1}{2}\ln \frac{L_{2-}}{r_0}
\ee

The divergence at small $L_{2-}$ in cut-off regularization is then
\be
S_{div}=-\left[\frac{R^2}{2\pi \alpha '}\frac{\sqrt{2}L_-}{r_0}\right]\frac{1}{4}\ln\epsilon\label{divcut}
\ee

Note however that this is the divergence at any value of $L_{2-}$. Indeed, the divergence comes from the 
lower limit of the integral. A simple calculation of the integral near the lower end shows that the divergence is 
in general
\be
S_{div}=\left[\frac{R^2}{2\pi \alpha '}\frac{\sqrt{2}L_-}{r_0}\right]\int_{1+\epsilon}\frac{dz}{4(z-1)}
=-\left[\frac{R^2}{2\pi \alpha '}\frac{\sqrt{2}L_-}{r_0}\right]\frac{1}{4}\ln\epsilon
\ee

\begin{figure}[bthp]
\begin{center}\includegraphics{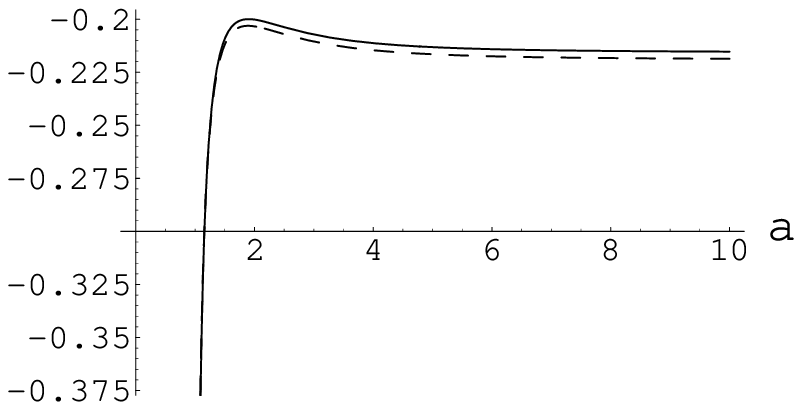}\end{center}
\caption{Integral in S, times $1/\sqrt{a^4+1}$, as a function of $a$, 
with the divergence subtracted, for $\epsilon=0.01$ (solid line) and
$\epsilon=0.001$ (dashed line)}\label{cutoffact}
\end{figure}

\begin{figure}[bthp]
\begin{center}\includegraphics{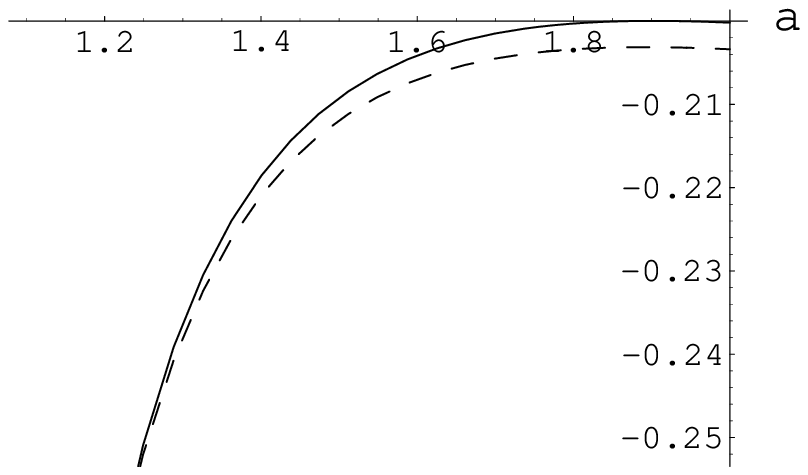}\end{center}
\caption{Integral in S, times $1/\sqrt{a^4+1}$, as a function of $a$, 
with the divergence subtracted, for $\epsilon=0.01$ (solid line) and
$\epsilon=0.001$ (dashed line)}\label{cutoffact2}
\end{figure}

Subtracting this divergence from the action and numerically evaluating and 
plotting the result we obtain the graph in Fig.\ref{cutoffact}
and Fig.\ref{cutoffact2} (we have also checked that this is the true divergence, down to $\epsilon=10^{-9}$, after
which numerical evaluation errors become important).

\vspace{2cm}


{\bf Dimensional regularization}

Another regularization one can take is dimensional regularization. At $T=0$, the natural choice in the CFT is 
dimensional reduction, i.e. dimensionally reducing 10d SYM to $d=4-2\epsilon$. Following Alday and Maldacena, this 
corresponds to taking a $p=3-2\epsilon$ brane in 10 dimensions (where $d=p+1$). 

Therefore, the naive extension of this choice to $T\neq 0$ would be to take the black 
near-extremal $p=3-2\epsilon$ brane in 
10 dimensions, with metric 
\be
ds^2=H_p^{-1/2}(-dt^2H+d\vec{x}_{d-1}^2)+H_p^{1/2}(\frac{dr^2}{H}+r^2d\Omega_{5+2\epsilon})
\ee 

This matches what we used at $\epsilon=0$. The near horizon limit of this near-extremal 
brane gives
\be
H_p=\frac{c_d\lambda_d}{r^{4+2\epsilon}};\;\;\;
H=1-\frac{r_0^{7-p}}{r^{7-p}}=1-\frac{r_0^{4+2\epsilon}}{r^{4+2\epsilon}}
\ee
After the T duality we  obtain the metric
\be
ds^2=\frac{\sqrt{c_d\lambda_d}}{r^{2+\epsilon}}(-\frac{dy_0^2}{1-(r_0/r)^{4+2\epsilon}}
+d\vec{y}_{d-1}^2+\frac{dr^2}{1-(r_0/r)^{4+2\epsilon}})\label{tdume}
\ee

We can easily see however that using this metric does not regulate the divergence in (\ref{tdact2}), since 
it comes from the $r-r_0$ factors in the metric, which are unmodified ($1-(r_0/r)^{4+2\epsilon}\simeq (4+2\epsilon)
(r-r_0)$ near $r_0$, as before). 

The point is that this dimensional regularization still regularizes $r=0$, not $r=r_0$. We see in fact that 
we have the overall $r^{-\epsilon}$ power needed to match the dimension of the coupling
\be
\lambda_d=\lambda \mu^{2\epsilon}(4\pi e^{-\gamma})^{-\epsilon};
\;\;\;\; \lambda=\frac{R^4}{{\alpha '} ^2}
\ee
except that unlike at $T=0$, the IR is not at $r=0$ anymore, but at $r=r_0$, whereas $\mu$ is supposed to be 
still an IR scale, which is not consistent. 
We need therefore to replace $r^{-\epsilon}$ with $(r-r_0)^{-\epsilon}$, or equivalently, 
to replace $\mu$ with $\mu r^{\epsilon}/(r-r_0)^{\epsilon}$, so that the redefined $\mu$ is again an IR scale.
We then obtain the metric 
\be
ds^2=\frac{\sqrt{c_d\lambda_d}}{r^{2}(r-r_0)^{\epsilon}}(-\frac{dy_0^2}{1-(r_0/r)^{4+2\epsilon}}
+d\vec{y}_{d-1}^2+\frac{dr^2}{1-(r_0/r)^{4+2\epsilon}})\label{newregmetric}
\ee

This will indeed regulate the divergence in (\ref{tdact2}). This seems like an ad-hoc procedure, especially since 
the new metric will not be a 10d supergravity solution anymore, but we will justify it a posteriori by matching 
with the results of cut-off regularization. 

Note that (\ref{wittenm}) (the Witten construction for the $AdS_{n+1}$ black hole in the $M\rightarrow \infty$
limit) gives a different dimensional regularization than the near horizon near-extremal $p$-brane in (\ref{tdume}),
but still doesn't regulate the divergence. The point is that $AdS_{n+1}$ is the near horizon of the extremal 
$n-1$ brane only if $n=4$, so it is not suited for us.

Using the metric in (\ref{newregmetric}) we get the action 
\be
S=\frac{\sqrt{\tilde{c}_d}}
{2\pi \alpha '}\frac{L_-}{\sqrt{2}}2\int_0^{L_{2-}/(2\sqrt{2})}dy_2\frac{R^2r_0^{2+\epsilon}r^\epsilon
\mu^{\epsilon}}{(r-r_0)^\epsilon(r^{4+2\epsilon}-r_0^{4+2\epsilon})}
\sqrt{r'^2+\frac{r^{4+2\epsilon}+r_0^{4+2\epsilon}}{r_0^{4+2\epsilon}r^{4+2\epsilon}}
(r^{4+2\epsilon}-r_0^{4+2\epsilon})}
\ee
where
\be
\sqrt{\tilde{c}_d}=\sqrt{c_d}(4\pi e^{-\gamma})^{-\epsilon/2}=(2\pi e^{\gamma/2})^\epsilon
(\Gamma(2+\epsilon))^{1/2}
\ee
Its equations of motion are 
\be
r'^2=\frac{r^{4+2\epsilon}+r_0^{4+2\epsilon}}{r_0^{4+2\epsilon}r^{8+4\epsilon}}
[\mu^{2\epsilon}R^4\alpha^2\frac{r^{2\epsilon}}{(r-r_0)^{2\epsilon}}
(r^{4+2\epsilon}+r_0^{4+2\epsilon})-r^{4+2\epsilon}
(r^{4+2\epsilon}-r_0^{4+2\epsilon})]
\ee
Taking into account that 
\be
\frac{\alpha R^2}{r_0^2}\frac{\mu^\epsilon}{r_0^{\epsilon}}=a^2(a-1)^{\epsilon}
\sqrt{\frac{a^{4+2\epsilon}-1}{a^{4+2\epsilon}+1}}
\ee
we obtain the regularized version of (\ref{tdact2})
\be
S=\frac{ R^2}{2\pi \alpha '}\frac{\sqrt{2}L_-\sqrt{\tilde{c}_d}\mu^{\epsilon}}
{r_0^{1+\epsilon}}\frac{(a-1)^{\epsilon}}{a^{\epsilon}\sqrt{a^{4+2\epsilon}+1}}
\int_1^a\frac{dz\; z^{2\epsilon}}{(z-1)^{2\epsilon}(z^{4+2\epsilon}-1)}
\sqrt{\frac{z^{4+2\epsilon}+1}{\frac{z^{4+2\epsilon}+1}{a^{4+2\epsilon}+1}
\frac{z^{2\epsilon}(a-1)^{2\epsilon}}{(z-1)^{2\epsilon}a^{2\epsilon}}-\frac{z^{4+
2\epsilon}}{a^{4+2\epsilon}}\frac{z^{4+2\epsilon}-1}{a^{4+2\epsilon}-1}}}\label{intact}
\ee

One could also write a regularized version of $L_{2-}$,
\be
\frac{L_{2-}}{2\sqrt{2}}=\frac{r_0}{a^{2+\epsilon}\sqrt{a^{4+2\epsilon}-1}}\int_1^{a}
\frac{dz z^{4+2\epsilon}}{\sqrt{z^{4+2\epsilon}+1}
\sqrt{\frac{z^{4+2\epsilon}+1}{a^{4+2\epsilon}+1}
\frac{z^{2\epsilon}(a-1)^{2\epsilon}}{(z-1)^{2\epsilon}a^{2\epsilon}}
-\frac{z^{4+2\epsilon}}{a^{4+2\epsilon}}
\frac{z^{4+2\epsilon}-1}{a^{4+2\epsilon}-1}}}
\ee
but we have seen that $L_{2-}$ is finite, so as long as the divergence is $a$-independent
(as we will find), we don't need the finite $\epsilon$ expression for $L_{2-}$.

We can calculate the leading divergence of the above action. We can check that at $\epsilon=0$
the integral is divergent at the lower end, as $\int_1 dz/(z-1)$, whereas at the upper end it is finite, since near 
the upper end it gives $\int^a dz/\sqrt{z-a}\propto \sqrt{z-a}|^a=0$. 

When we introduce the epsilon regularization, things are a bit more subtle, and the integral resembles 
\be
\int_0^a\frac{dx}{x^{1+\epsilon}}=\frac{x^{-\epsilon}}{-\epsilon}|_0^a=\frac{a^{-\epsilon}}{-\epsilon}
\simeq -\frac{1}{\epsilon}+\ln a 
\ee
if $\epsilon <0$ and it seems as if there is no contribution from the lower end $x=0$ after introducing $\epsilon$.
But this is not quite true, since one can split the region of integration with a $\delta$, $1\gg \delta\gg \epsilon$
and then 
\be
\int_0^a\frac{dx}{x^{1+\epsilon}}=(\int_0^\delta+\int_\delta^a)\frac{dx}{x^{1+\epsilon}}=\frac{x^{-\epsilon}}{-\epsilon
}|_\delta ^a+\frac{x^{-\epsilon}}{-\epsilon}|_0^\delta=(\ln a-\ln \delta)+(-\frac{1}{\epsilon}+\ln \delta)
\ee
so we see that the $1/\epsilon$ divergence actually still comes from around $x=0$. In fact, our integral is similar 
to the above if we have a finite function $f(x)$ inserted in the integral. Looking near $x=0$ we get the correct 
divergence $-f(0)/\epsilon$. 

Therefore the divergent part of the integral in (\ref{intact}) is given by 
\be
I\simeq \frac{\sqrt{a^4+1}}{4}\int_1\frac{dz}{(z-1)^{1+\epsilon}}\simeq -\frac{\sqrt{a^4+1}}{4\epsilon}+...
\ee
The finite part of this integral is not relevant since it will be different than the finite part of the 
true integral. Replacing this divergence in the action, we get 
\bea
&&S=\frac{R^2}{2\pi \alpha '}\frac{\sqrt{2}L_-}{r_0}\sqrt{\tilde{c}_d}\left(\frac{\mu}{r_0}\right)^\epsilon
[-\frac{1}{\sqrt{a^4+1}}\frac{\sqrt{
a^4+1}}{4\epsilon}+f(a)]\nonumber\\&&
=\frac{R^2}{2\pi \alpha '}\frac{\sqrt{2}L_-}{r_0}[-\frac{1}{4\epsilon}-\frac{1}{4}\ln\frac{\mu}{
r_0}-\frac{\ln (2\pi \sqrt{e})}{4}+ f(a)]
\eea
where $a=a(L_{2-}/r_0)$ and $f(a)$ is the finite part of the integral (with extra terms coming also from the a-dependent
prefactor of the integral), that can be determined numerically.

We now compare the divergence with the cut-off regularization case. When doing this, we need to drop the 
$\epsilon_{dim.reg.}$ term, and compare the $\ln \mu/r_0$ term with the divergence of cut-off regularization
in (\ref{divcut}). We see that they match if 
\be
\frac{\mu}{r_0}=\epsilon_{cut-off}=\frac{\Lambda}{r_0}
\ee
therefore the modified dimensional regularization scale $\mu$ equals the cut-off in the $r$ integration, $\Lambda$.

This justifies a posteriori the choice of dimensional regularization metric in (\ref{newregmetric}), as promised.
Notice that in the identification of $\mu$ with $\Lambda$ there could be a priori a constant, $\mu=c\Lambda$, 
which will change the identification of the finite parts by a constant.

\begin{figure}[bthp]
\begin{center}\includegraphics{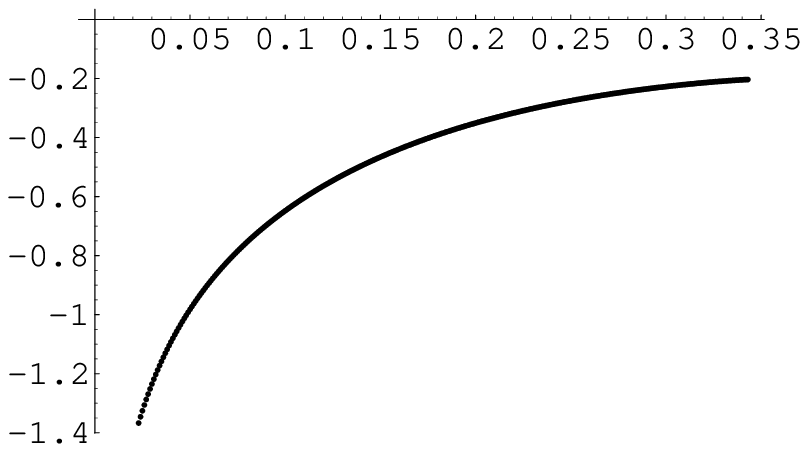}\end{center}
\caption{Finite part of the integral in S, times $1/\sqrt{a^4+1}$, as function of $L_{2-}/(2\sqrt{2}r_0)$}\label{sofl}
\end{figure}

As for the finite part of (\ref{intact}), 
it is composed of two contributions. Since the divergence comes from the lower end of the 
integration, the bulk of the integration is finite, and to calculate it we can put $\epsilon=0$, therefore 
obtaining just the finite part of the cut-off regularization. There is one more contribution coming from the 
subleading part (in $\epsilon$) of the divergence near $z=1$. Expanding the integral 
and the $a$-dependent prefactor in $\epsilon$ near $z=1$, 
we get 
\bea
&&S_{div}=\frac{ R^2}{2\pi \alpha '}\frac{\sqrt{2}L_-\sqrt{\tilde{c}_d}\mu^{\epsilon}}
{r_0^{1+\epsilon}}\int_1\frac{dz}{(z-1)^{1+\epsilon}}\frac{z^\epsilon}{4+2\epsilon}\left[1-(z-1)^{1+2\epsilon}
\frac{4+2\epsilon}{2+(z-1)(4+2\epsilon)}\right.\nonumber\\&&\left.\times
z^4\frac{a^{4+2\epsilon}+1}{a^4(a^{4+2\epsilon}-1)(a-1)^{2\epsilon}}\right]
^{-1/2}
\eea
but now further expanding in $z-1$ the $z^{\epsilon}$ and $[...]^{-1/2}$ will give no new finite contribution, since 
\be
\int_1^\delta \frac{dz}{(z-1)^{1+\epsilon}}(z-1)^{1+2\epsilon}=\frac{(z-1)^{1+\epsilon}}{1+\epsilon}|_1^\delta=\frac{\delta^{1+
\epsilon}}{1+\epsilon}\rightarrow 0
\ee
Therefore the action is 
\bea
&&S=\frac{ R^2}{2\pi \alpha '}\frac{\sqrt{2}L_-}
{r_0}\left[\left(\frac{\mu}{r_0}\right)^\epsilon\frac{\sqrt{\tilde{c}_d}}
{4+2\epsilon}\int_1\frac{dz}{(z-1)^{1+\epsilon}}
+\tilde{f}(a)\right]\nonumber\\&&
=\frac{R^2}{2\pi \alpha '}\frac{\sqrt{2}L_-}{r_0}\left[-\frac{1}{4\epsilon}-\frac{1}{4}\ln\frac{2\pi \mu}{
r_0} +\tilde{f}(a)\right]
\eea
where $\tilde{f}(a)=f(a)-1/8$ is the same finite part as in the cut-off regularization, except for a possible constant 
term, due to the fact that we could actually have $\mu=c\Lambda$. We have numerically checked that if $\mu=\Lambda$,
$\tilde{f}(a)=$ the finite part in cut-off regularization, to an accuracy of $<10^{-6}$. 
Otherwise, the difference 
in finite parts in the two regularization is just a number, and we are only interested in the $a$ dependence, which 
is still the one plotted in Fig.\ref{cutoffact} and Fig.\ref{cutoffact2}. Therefore we can also say that the errors
on Fig.\ref{cutoffact},\ref{cutoffact2} are $<10^{-6}$. Finally, we can substitute the $a(L_{2-}/r_0)$ dependence 
inside $S_{finite}(a)$ and obtain the graph for $S_{finite}(L_{2-}/r_0)$ in Fig.\ref{sofl}.

Let us now understand this result in ${\cal N}=4$ SYM. Since $R^2/\alpha '=\sqrt{\lambda}$ and $r_0=TR^2/\pi$,
\be
\ln a_4=iS=i\frac{L_-/\alpha '}{\sqrt{2}T}
\left[-\frac{1}{4\epsilon}-\frac{1}{4}\ln\frac{2\pi^2}{
\sqrt{\lambda}}\frac{\mu/\alpha '}{T} +\tilde{f}(a)\right]
\ee
and 
\be
a=a\left(\frac{\pi}{\sqrt{\lambda}}\frac{L_{2-}/\alpha '}{T}\right)
\ee
It is quite different from the $T=0$ result (\ref{amplres}), especially in that it depends on $L_{2-}/T,L_{-}/T$
and $\mu/T$ instead of $L_{2-}L_-/\mu^2$. One would have had $f_0(\lambda)=\sqrt{\lambda}/\pi$ 
(the zero temperature cusp anomalous dimension) in front, but $r_0$ in the denominator
contained also $\sqrt{\lambda}$. Note however that the pure phase part of (\ref{amplres}) is also a single log, 
as is the case now, but $\ln (L_-L_{2-})/\mu^2$ is replaced with $(L_-/T)\ln (\mu/T)$ + finite ($(L_-/T)\ln (L_{2-})/T$ 
at small $L_{2-}$).

The issue of regularization deserves some comments. For the Wilson loop, at least in the static case (for 
the calculation of $V_{q\bar{q}}(L)$) in the gravity dual one subtracts the divergence in the area of the 
string worldsheet and interprets it as due to the mass of the W bosons (divergence = perimeter of contour $C$ $\times$
length of string from contour at $r=\infty$ to $r_1$, see the Appendix). In the case of the gluon amplitude 
calculation, the worldsheet defined in the T dual metric is also divergent. But the divergence is interpreted, 
as we saw, as a field theory divergence. In the case of cut-off regularization, the divergence in $\ln r_0/\Lambda$
(that for the Wilson loop case would be exactly the W boson mass divergence) corresponds in field theory
to an IR cut-off divergence. In dimensional regularization we also have the $1/\epsilon$ divergence at $d=4-2\epsilon$
on top of it.

\section{Conclusions}

In this paper we have generalized the construction of Alday and Maldacena \cite{am} for gluon scattering in 
${\cal N}=4$ Super Yang-Mills from AdS-CFT to finite temperature. At $T=0$, after a "T duality" the space 
becomes again AdS, and the gluon amplitude is $A_n=A_{n,tree}e^{iS_{string}(C)}$, where the string worldsheet 
ends on a lightlike polygon at the UV boundary of the T dual metric. In the original metric, the polygon is 
situated in the IR, but still on the boundary of space, because of the limit involved. The calculation of 
$e^{iS_{string}(C)}$ gives the same result as for the Wilson loop.

At nonzero $T$, the gravity dual is the Poincar\'{e} AdS black hole metric, but now the T duality produces a 
different metric. As a result, the calculation of $e^{iS_{string}}$ for the T dual metric (for gluon 
scattering) produces a different result than the 
calculation of the Wilson loop. We have studied the possible string worldsheets in the case of a lightlike 
polygon of a particular type, related to the usual $V_{q\bar{q}}(L)$ contour by making the sides lightlike.
These should all correspond to observables in ${\cal N}=4$ SYM. 
We have found that the possible string worldsheets are the usual one, ending in the UV of the original metric
and giving the lighlike Wilson loop, the new one, ending at the black hole horizon, but in the T dual metric, 
and defining gluon scattering, and a strange third case, defined close to the horizon in the original metric.
These worldsheets now give different ${\cal N}=4$ SYM observables at finite temperature. The Wilson loop is 
related by a boost to the Wilson loop that was argued to give the jet quenching parameter as in \cite{lrw}
(see \cite{aev} though).

We have argued that gluon scattering is still defined by $A_n=A_{n,tree}e^{iS_{string}(C)}$, with the 
worldsheet defined in the T dual metric and ending at the horizon. In the original metric, it ends on the 
black hole horizon, but we have shown that like in the $T=0$ case, even though this is the IR of the metric, 
it is on the boundary because of the limits involved. The gluon amplitudes at finite temperature are still 
IR divergent (for on-shell, massless, external gluons), and depend on more parameters than at $T=0$, since 
the heat bath breaks Lorentz invariance. For the particular amplitude studied, instead of dependence 
on $s/\mu^2=-t/\mu^2=L_-L_{2-}/(2\pi \mu\alpha ')^2$, we found dependence on $L_-/T$, $L_{2-}/T$ and $\mu/T$.
Therefore we have argued that the Wilson loop observable still defines a gluon scattering amplitude.

We note that there is a potential problem with the definition of the gluon external states at finite $T$ and 
strong coupling. However, in the gravity dual the observable is well defined, thus a good definition of the gluon 
states should exist, but we have not explored this further. It also helps that in the particular case we examined, 
one gluon has energy $E\gg T$ (so for it the temperature can be treated as a small irrelevant perturbation). 

We have studied the 4-point gluon amplitude defined by a lightlike polygon with rectangular spatial
projection and $L_-\gg L_{2-}$ and colour ordering in, in, out, out. It corresponds to forward scattering 
of an low energy gluon ($E<T$) off an energetic gluon ($E\gg T$). We have started with the condition $L_-\gg 
L_{2-}$ only, but we have found that $L_{2-}/r_0$ is bounded by $\simeq 0.35$, implying $E<T$ for the low 
energy gluon. In particular that means that there is no $T\rightarrow 0$ limit of this calculation, which is 
good, since the $T=0$ result (\ref{amplres}) contains both a real and an imaginary part, which could not 
come from a real worldsheet. 
This amplitude should be related in some way to the jet quenching of gluon jets in the ${\cal N}=4
$ SYM plasma, however the relation is not immediately obvious. As we saw, the jet quenching parameter (for 
quarks) is related to the Wilson loop, which gives a different result. At small $L$, the partially light-like
Wilson loop gives $e^{iS}\propto e^{-a T^3 L_-L^2}$ with $a$ a number, whereas now we obtain $e^{iS}\propto
e^{ia'(L_-/(\alpha 'T))\ln(L_{2-}/(\alpha 'T))}$. If we would know the 4-point amplitude at arbitrary values of momenta, we could 
calculate the viscosity coefficient using (\ref{visco}).

The above 4-point amplitude was calculated using cut-off and generalized dimensional regularization. 
The cut-off regularization involves integrating only up to $r_0+\Lambda =r_0(1+\epsilon)$ in the gravity 
dual, and corresponds in field theory to straightforward IR cut-off regularization. Straightforward dimensional 
regularization following Alday and Maldacena is not possible at $T=0$ since it does not regulate the IR 
divergence, which is now in $r-r_0$, not in $r$. We needed therefore to modify $\mu\rightarrow \mu r^\epsilon/
(r-r_0)^\epsilon$ in order to regulate the divergence. This does not correspond to a solution of the supergravity 
equations of motion anymore, so it is a bit ad-hoc, but we justified it a posteriori by matching with the 
cut-off regularization. The choice $\Lambda =\mu$ matches the divergencies in the two regularizations, and the 
finite parts are also found to be related. This interpretation of the divergencies for the gluon amplitudes is 
to be contrasted with the Wilson loop calculation, for which the same $r\rightarrow r_0$ divergence is interpreted
as an infinite mass of W bosons, and is subtracted to give the Wilson loop.

Finally, it should be interesting to calculate the 4-point gluon amplitude at arbitrary values of the momenta, 
and relate it to the viscosity. But the computation of the string worldsheet seems quite challenging in the 
general case.

{\bf Acknowledgements}

We thank S. Dobashi for useful discussions.
K. Ito is supported in part by Ministry of Eduation, Culture, Sports, Science
and Technology of Japan.
The research of H.N. has been done with partial support from MEXT's program
"Promotion of Environmental Improvement for Independence of Young Researchers"
under the Special Coordination Funds for Promoting Science and Technology. 

\newpage

\section{Appendix}

\subsection{Wilson loops for $V_{q\bar{q}}(L)$}

For completeness, we quickly review how one calculates the worldsheet of the strings giving $V_{q\bar{q}}(L)$ 
at $T=0$ and finite $T$, and also derive the worldsheet ending on the same contour in the T dual AdS black 
hole metric (\ref{tdualmetric}).

At $T=0$, the calculation was done in \cite{maldacena2,ry}. The AdS metric is 
\be
ds^2=\frac{r^2}{R^2}(-dt^2+d\vec{x}^2)+R^2\frac{dr^2}{r^2}
\ee
and we consider the Wilson loop contour for the static $q\bar{q}$ potential (as in fig.\ref{contours}a) to be defined
at $r\rightarrow \infty$ and a string worldsheet ending on it. In the limit $T\gg L$ (here $T=$ time, and 
corresponds to the long side of the Wilson loop), we have a time translationally invariant worldsheet. Thus 
we can choose the static gauge $\tau=t,\sigma= x_2$ and the ansatz $r=r(\sigma)=r(x_2)$. Then we get the action
\be
S\propto\int dx_2\sqrt{r'^2+\frac{r^4}{R^4}}
\ee

Since this Lagrangian density is independent of $x_2$ the Hamiltonian (with $y_2$= "time") is independent of
$y_2$ and equal to $\equiv \alpha$. We then obtain the equation of motion
\be
r'^2=\frac{r^4}{R^4\alpha^2}(\frac{r^4}{R^4}-\alpha^2)
\ee

This means that the turning point of the worldsheet $r'=0$ is situated at $r=R\sqrt{\alpha}$, and for $r$ to be 
real we need that $r>R\sqrt{\alpha}$, i.e. we must put the Wilson loop contour (boundary of the worldsheet)
at infinity.

For the finite temperature case, the metric is 
\be
ds^2=\frac{r^2}{R^2}(-h dt^2+d\vec{x}^2)+R^2\frac{dr^2}{h r^2}
\ee

Using the same static gauge $\tau=t,\sigma= x_2$ and ansatz $r=r(\sigma)=r(x_2)$, we obtain the action
\be
S\propto
\int dx_2\sqrt{r'^2+\frac{r^4-r_0^4}{R^4}}
\ee

Again the Lagrangian density is independent of $x_2$, thus the Hamiltonian is constant and $\equiv \alpha$, 
giving the equation of motion 
\be
r'^2=\frac{r^4-r_0^4}{R^4\alpha^2}(\frac{r^4-r_0^4}{R^4}-\alpha^2)=\frac{r^4-r_0^4}{R^8\alpha^2}
(r^4-r_1^4)
\ee
where 
\be
r_1=r_0(1+\alpha^2R^4/r_0^4)^{1/4}
\ee

The turning point $r'=0$ seems to be either $r=r_1$ or $r=r_0$, however we need $r\geq r_1$ to have $r$ real 
($r'^2\geq 0$), therefore we can again only define the Wilson loop contour at $r=\infty$ and then the string 
stretches down to $r=r_1(\alpha)$. As $L$ increases, $r_1(\alpha)\rightarrow r_0$, and the string stretches for a
longer and longer period almost parallel to the horizon ($r\simeq r_0$), which is the reason we get the area law 
for the Wilson loop.

For the T dual AdS black hole, the metric is 
\be
ds^2=\frac{R^2}{r^2}(-\frac{ dt^2}{h}+d\vec{x}^2)+R^2\frac{dr^2}{h r^2}
\ee

Using the same static gauge $\tau=t,\sigma= x_2$ and ansatz $r=r(\sigma)=r(x_2)$, we obtain the action
\be
S\propto\int dy_2\frac{R^2}{r^2-r_0^4/r^2}\sqrt{r'^2+1-\frac{r_0^4}{r^4}}
\ee

If the Hamiltonian is $=\alpha$, we obtain the equations of motion 
\be
r'^2=\frac{1}{r^4}[\frac{R^4}{\alpha^2}+r_0^4-r^4]=\frac{1}{r^4}[r_1^4-r^4]
\ee
where $r_1^4=r_0^4+R^4/\alpha^2$. This means that we have a real $r$, i.e. $r'^2\geq 0$ only if $r\leq r_1$. 
Therefore in this case we have a Wilson loop contour defined at the horizon $r_0$, and a string that stretches 
from it to $r_1(\alpha)$.

\subsection{Penrose diagram of the Poincar\'{e} AdS black hole}

In this Appendix we will derive the Penrose diagram of the Poincar\'{e} AdS black hole, in order to see if the 
endpoints of our open string worldsheet are on the boundary of this gravity dual. The metric (using the Witten
construction as in (\ref{wittenm})) is
\be
ds^2=\frac{r^2}{R^2}[-dt^2(1-\frac{r_0^n}{r^n})+d\vec{y}_{(n-1)}^2]+R^2\frac{dr^2}{r^2(1-r_0^n/r^n)}
\ee
As noted in the text, this is a certain limit of the usual global AdS black hole, so we need to find its Penrose 
diagram from the begining.

{\em Warm up: the Schwarzschild and the $AdS_2$ black hole case}

It seems that we need to analyze 3 coordinates ($t,x,r$) to understand
this Penrose diagram. So we will simplify the analysis and look first at the case of the $AdS_2$ black hole, 
by getting rid of the 3 $x_i$ coordinates. The $AdS_2$ black hole will be (defining $r/R\equiv R/\tilde{r}$)
\be
ds^2=\frac{R^2}{\tilde{r}^2}\left(-dt^2(1-\frac{\tilde{r}}{\tilde{r}_0})+\frac{d\tilde{r}^2}{
1-\frac{\tilde{r}}{\tilde{r}_0}}\right)
\ee
which looks similar to the 4d Schwarzschild black hole reduced over the angles (except for the 
overall conformal factor and the fact that we have  $\tilde{r}/\tilde{r}_0$ instead of $r_0/r$), so we will 
try the same kind of coordinate transformations as for the Schwarzschild case. Let us review it first.

The metric of the 4d Schwarzschild black hole reduced over $\Omega_2$ is 
\be
ds^2=-dt^2(1-\frac{r_0}{r})+\frac{dr^2}{1-\frac{r_0}{r}}
\ee

The first transformation one does satisfies $dr=(1-r_0/r)dr_*$ and is
\be
r_*=r+r_0\ln (\frac{r}{r_0}-1)
\ee

Then one changes to lightcone coordinates, $u=t-r_*,v=t+r_*$ followed by the change to Kruskal coordinates
\be
\bar{u}=-2r_0e^{-u/(2r_0)};\;\;\; \bar{v}=2r_0e^{v/(2r_0)}
\ee
such that 
\be
dudv=d\bar{u}d\bar{v}e^{\frac{u-v}{2r_0}}=d\bar{u}d\bar{v}\frac{e^{-r/r_0}}{r/r_0-1}
\ee
i.e., so that  one can eliminate the singular metric factor.
We then get 
\be
ds^2=-\frac{r_0}{r}e^{-r/r_0}d\bar{u}d\bar{v}
\ee

This means that $\bar{u}<0,\bar{v}>0$, but one can analytically extend to the whole $(\bar{u},\bar{v})$ plane, 
thus getting the Kruskal diagram. Finally, one changes variables from the 2d Minkowski space in $\bar{u},\bar{v}$
to the coordinates of its Penrose diagram.

To get the Penrose diagram of the fully extended black hole, 
one next analyzes the position of the boundaries, singularities and horizons in the diagram.
Now the boundary at infinity $r=\infty$ becomes $r_*=+\infty$, or $\bar{u}=-\infty,\bar{v}=+\infty$. The horizon 
$r=r_0$ corresponds to $r_*=-\infty$, or $\bar{u}=\bar{v}=0$.   Since 
\be
\bar{u}\bar{v}=-4r_0^2e^{r/r_0}(r/r_0-1)
\ee
the horizon $r=r_0$ is now at $\bar{u}\bar{v}=0$ and the singularity behind it, $r=0$, is at 
$\bar{u}\bar{v}=4r_0^2$.

We now apply the same logic to the $AdS_2$ black hole case. We first make a transformations that satisfies
$d\tilde{r}=(1-\tilde{r}/\tilde{r}_0)dr_*$, namely
\be
r_*=-\tilde{r}_0\ln (1-\frac{\tilde{r}}{\tilde{r}_0})
\ee
Since we have $\tilde{r}\leq \tilde{r}_0$, the boundary $\tilde{r}=0$ becomes $r_*=0$ and 
the horizon $\tilde{r}=\tilde{r}_0$ becomes $r_*=\infty$. Then we change to lightcone 
coordinates $u=t-r_*,v=t+r_*$ and make the transformation
\be
\bar{u}=2\tilde{r}_0e^{u/(2\tilde{r}_0)};\;\;\;
\bar{v}=-2\tilde{r}_0e^{-v/(2\tilde{r}_0)}
\ee
which eliminates the singular factor in the metric, since 
\be
dudv=d\bar{u}d\bar{v}\frac{1}{1-\tilde{r}/\tilde{r}_0}
\ee
giving the metric in Kruskal-like coordinates
\be
ds^2=-\frac{R^2}{\tilde{r}^2}d\bar{u}d\bar{v}
\ee

This is again nonsingular at the horizon $\tilde{r}=\tilde{r}_0$, as for the Schwarzschild case,
so we can analytically
continue through it. Note that now the metric looks similar to the pure 
$AdS_2$ metric in Poincar\'{e}
coordinates, except that now $\tilde{r}=\tilde{r}_0(1+\bar{u}\bar{v}/(4\tilde{r}_0^2))$
instead of $\tilde{r}=(\bar{v}-\bar{u})/2$ for $AdS_2$. 

To complete the Kruskal-like diagram we need the position of the boundary, horizon and singularity.
We have
\be
\bar{u}\bar{v}=-4\tilde{r}_0^2(1-\frac{\tilde{r}}{\tilde{r}_0})
\ee
and the original region $0\leq \tilde{r}\leq \tilde{r}_0$ is defined by $0\geq \bar{u}\bar{v}\geq -4\tilde{r}_0^2$ and 
$\bar{u}\geq 0,\bar{v}\leq 0$.

The boundary $\tilde{r}=0$ is now at $\bar{u}\bar{v}=-4\tilde{r}_0^2$, the horizon $\tilde{r}
=\tilde{r}_0$ is at $\bar{u}\bar{v}=0$ and the singularity $\tilde{r}=\infty$ is at $
\bar{u}\bar{v}=+\infty$. 

Finally, in order to find the Penrose diagram, we make the usual
coordinate transformation of flat 2 dimensional Minkowski space in $\bar{u},\bar{v}$ to 2d $\tau,\theta$
coordinates of finite extent. Indeed, the usual transformation
\be
\bar{u}=2\tilde{r}_0\tan \tilde{u};\;\;\; 
\bar{v}=2\tilde{r}_0\tan \tilde{v};\;\;\;
\tilde{u}=\tau -\theta;\;\;\;
\tilde{v}=\tau +\theta
\ee
defines as usual the Penrose diagram in $\tau, \theta$, giving the metric
\be
ds^2=-\frac{R^2}{\tilde{r}^2}
d\bar{u}d\bar{v}=\frac{4R^2}{\tilde{r}^2}
\frac{\tilde{r}_0^2}{\cos^2\tilde{u}\cos^2\tilde{v}}(-d\tau^2+d\theta^2)
\ee

\begin{figure}[bthp]
\begin{center}\includegraphics{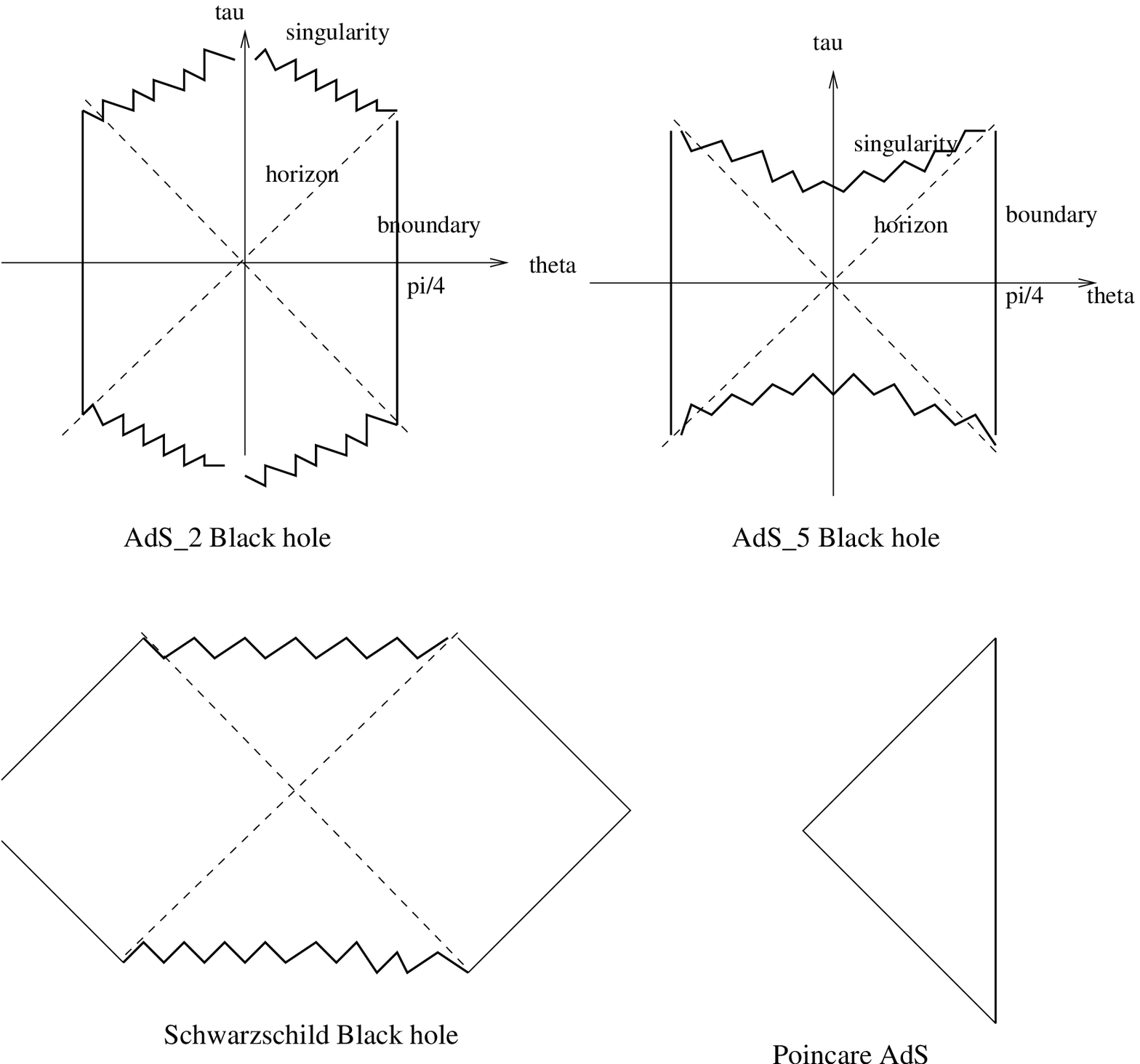}\end{center}
\caption{Penrose diagrams of: $AdS_2$ BH; $AdS_5 BH$; Schwarzschild BH; AdS in Poincar\'{e} 
coordinates}\label{adsbhpenrose}
\end{figure}

In these coordinates, the boundary  at $\bar{u}\bar{v}=-4\tilde{r}_0^2$ is now at $\theta=\pm \pi/4$,
the horizon at $\bar{u}\bar{v}=0$ is at $\tau=\pm\theta$ and the singularity at $\bar{u}\bar{v}=
+\infty$ is at $\tau\pm \theta=\pm\pi/2$. Thus the Penrose diagram for the maximal extension 
of the $AdS_2$ black hole in Poincar\'{e} coordinates is as in Fig.\ref{adsbhpenrose}.

As observed in the figure, the horizon touches the boundary, on the points of $t=\pm\infty$, which 
corresponds to $u=v=+\infty$; $\bar{u}=+\infty,\bar{v}= 0$; $\tau=-\theta=\pm \pi/4$ or $\tau=+\theta=
\pm \pi/4$. 

Thus as argued, the horizon touches the boundary on $t=\infty$ for the 2 dimensional Poincar\'{e} AdS black hole. 

{\em The $AdS_5$ black hole case}

For the first coordinate transformation we now need 
$d\tilde{r}=dr_*(1-\tilde{r}^4/\tilde{r}_0^4)$, thus
\be
r_*=\frac{\tilde{r}_0}{2}Arctan\frac{\tilde{r}}{\tilde{r}_0}+\frac{\tilde{r}_0}{4}\ln
\frac{\tilde{r}_0+\tilde{r}}{\tilde{r}_0-\tilde{r}}
\ee
giving the metric 
\be
ds^2=\frac{R^2}{\tilde{r}^2}(1-\frac{\tilde{r}^4}{\tilde{r}_0^4})[-dt^2+dr_*^2+\frac{
d\vec{x}_3^2}{1-\tilde{r}^4/\tilde{r}_0^4}]
\ee
After changing to lightcone coordinates $u=t-r_*,v=t+r_*$, we make the coordinate transformation
to Kruskal-like coordinates
\be
\bar{u}=\frac{\tilde{r}_0}{2}e^{2u/\tilde{r}_0};\;\;\;
\bar{v}=-\frac{\tilde{r}_0}{2}e^{-2v/\tilde{r}_0}
\ee

It eliminates the singular factor in the metric, since
\be
dudv=d\bar{u}d\bar{v}e^{\frac{4r_*}{\tilde{r}_0}}=d\bar{u}d\bar{v}e^{2Arctan(\tilde{r}/
\tilde{r}_0)}\frac{\tilde{r}_0+\tilde{r}}{\tilde{r}_0-\tilde{r}}
\ee
We then get 
\be
ds^2=\frac{R^2}{\tilde{r}^2}(1+\frac{\tilde{r}}{\tilde{r}_0})^2(1+\frac{\tilde{r}^2}{\tilde{r}
_0^2})[-e^{2Arctan (\tilde{r}/\tilde{r}_0)}d\bar{u}d\bar{v}+\frac{d\vec{x}_3^2}{
(1+\frac{\tilde{r}}{\tilde{r}_0})^2(1+\frac{\tilde{r}^2}{\tilde{r}
_0^2})}]
\ee
Finally, we transform to Penrose diagram coordinates as usual, by
\be
\bar{u}=\frac{\tilde{r}_0}{2}\tan \tilde{u};\;\;\;
\bar{v}=\frac{\tilde{r}_0}{2}\tan \tilde{v};\;\;\;
\tilde{u}=\tau-\theta;\;\;\;
\tilde{v}=\tau +\theta
\ee
giving
\be
ds^2=\frac{R^2}{\tilde{r}^2}(1+\frac{\tilde{r}}{\tilde{r}_0})^2(1+\frac{\tilde{r}^2}{\tilde{r}
_0^2})\frac{\tilde{r}_0^2e^{2Arctan (\tilde{r}/\tilde{r}_0)}}
{4\cos^2\tilde{u}\cos^2\tilde{v}}[-d\tau^2+d\theta^2+d\vec{x}_3^2
\frac{4e^{-2Arctan(\tilde{r}/\tilde{r}_0)}\cos^2\tilde{v}\cos^2\tilde{u}}{\tilde{r}_0^2
(1+\frac{\tilde{r}}{\tilde{r}_0})^2(1+\frac{\tilde{r}^2}{\tilde{r}
_0^2})}]
\ee

To complete the Penrose diagram, we need the position of the boundaries, horizons and singularities.
We first define them in Kruskal-like coordinates. We have
\be
\bar{u}\bar{v}=-\frac{\tilde{r}_0^2}{4}e^{-4r_*/\tilde{r}_0}=-\frac{\tilde{r}_0^2}{4}e^{-2
Arctan(\tilde{r}/\tilde{r}_0)}
\frac{1-\tilde{r}/\tilde{r}_0}{1+\tilde{r}/\tilde{r}_0}
\ee

Then the boundary  at $\tilde{r}=0$ becomes $\bar{u}\bar{v}=-\tilde{r}_0^2/4$, the 
horizon at $\tilde{r}=\tilde{r}_0$ becomes $\bar{u}\bar{v}=0$, and the singularity 
at $\tilde{r}=\infty$ becomes $\bar{u}\bar{v}=+\tilde{r}_0^2e^{-\pi}/4$.

Finally, in the Penrose diagram coordinates, the boundary is at $\theta=\pm \pi/4$, and the $d\vec{x}_3^2$ term 
in the square brackets remains finite (relative to $d\theta^2-d\tau^2$) if $\tilde{u}=
\tau-\pi/4$ and $\tilde{v}=\tau +\pi/4$ remain $\neq \pm \pi/2$. The horizon is at 
$\tau=\pm \theta$, and the $d\vec{x}_3^2$ term 
in the square brackets remains finite (relative to $d\theta^2-d\tau^2$) if $\theta\neq\pm\pi/4$. The singularity is at 
\be
\tan^2\tau=\frac{\tan^2\theta+e^{-\pi}}{e^{-\pi}\tan^2\theta+1}
\ee
and a similar comment applies. Then the Penrose diagram is as given in Fig.\ref{adsbhpenrose}.

Therefore the horizon and the boundary (and the singularity also) touch each other on 
the line of $\tau=\pm\theta=\pm \pi/4$, thus $\tilde{u}=0,\tilde{v}=\pi/2$ or $\tilde{v}
=0,\tilde{u}=\pi/2$, i.e. $\bar{u}=0,\bar{v}=-\infty$ or $\bar{v}=0,
\bar{u}=+\infty$, therefore $t=-\infty$ or $t=+\infty$. And in both cases, the $d\vec{x}_3^2$
term drops out ($\cos^2\tilde{u}\cos^2\tilde{v}=0$), hence $\vec{x}_3$ is arbitrary.

Therefore we have proven the assertion that the boundary touches the horizon on $t=\pm\infty$,
$x_i$ arbitrary.

\end{document}